\documentclass[aps, prx, a4paper, floatfix,  twocolumn, longbibliography]{revtex4-1}

\usepackage[utf8]{inputenc}
\usepackage{epsfig}
\usepackage[T1]{fontenc}
\usepackage[english]{babel}
\usepackage[table]{xcolor}
\usepackage{t1enc}
\usepackage{graphicx}
\usepackage{amssymb}
\usepackage{amsmath}
\usepackage{relsize}
\usepackage[percent]{overpic}
\usepackage{bm}
\usepackage{hyperref}
\usepackage[section]{placeins}

\usepackage[normalem]{ulem}

\newcommand{\avg}[1]{{\left<#1\right>}}

\newcommand{\dd}{\mathrm{d}}
\newcommand{\ee}{\mathrm{e}}

\usepackage{mathtools}
\def\multiset#1#2{\ensuremath{\left(\kern-.3em\left(\genfrac{}{}{0pt}{}{#1}{#2}\right)\kern-.3em\right)}}

\usepackage{amsmath}

\newcommand{\bb}{\bm{b}}
\newcommand{\A}{\bm{A}}
\newcommand{\D}{\bm{\mathcal{D}}}
\newcommand{\G}{\bm{G}}

\newcommand{\e}{\bm{e}}
\newcommand{\btau}{\bm{\tau}}

\usepackage{verbatim}
\usepackage{overpic}
\usepackage{booktabs}
\usepackage{placeins}

\usepackage{siunitx}
\usepackage{multirow}

\begin{document}

\title{Network reconstruction and community detection from dynamics}

\author{Tiago P. Peixoto}
\email{peixotot@ceu.edu}
\affiliation{Department of Network and Data Science, Central European University, H-1051 Budapest, Hungary}
\affiliation{ISI Foundation, Via Chisola 5, 10126 Torino, Italy}
\affiliation{Department of Mathematical Sciences, University of Bath, Claverton Down, Bath BA2
  7AY, United Kingdom}

\begin{abstract}
We present a scalable nonparametric Bayesian method to perform network
reconstruction from observed functional behavior that at the same time
infers the communities present in the network. We show that the joint
reconstruction with community detection has a synergistic effect, where
the edge correlations used to inform the existence of communities are
also inherently used to improve the accuracy of the reconstruction
which, in turn, can better inform the uncovering of communities. We
illustrate the use of our method with observations arising from epidemic
models and the Ising model, both on synthetic and empirical networks, as
well as on data containing only functional information.
\end{abstract}

\maketitle

The observed functional behavior of a wide variety large-scale system
is often the result of a network of pairwise interactions. However, in
many cases these interactions are hidden from us, either because they
are impossible to measure directly, or because their measurement can
be done only at significant experimental cost. Examples include the
mechanisms of gene and metabolic regulation~\cite{wang_inferring_2006},
brain connectivity~\cite{breakspear_dynamic_2017}, the spread of
epidemics~\cite{keeling_estimating_2002}, systemic risk in financial
institutions~\cite{musmeci_bootstrapping_2013}, and influence in social
media~\cite{bakshy_role_2012}. In such situations, we are required to
\emph{infer} the network of interactions from the observed functional
behavior. Researchers have approached this reconstruction
task from a variety of angles, resulting in many different methods,
including thresholding the correlation between
time-series~\cite{kramer_network_2009}, inversion of deterministic
dynamics~\cite{timme_revealing_2007,shandilya_inferring_2011,nitzan_revealing_2017},
statistical inference of graphical models~\cite{abbeel_learning_2006,
bresler_reconstruction_2008, montanari_which_2009,
hofling_estimation_2009, nguyen_inverse_2017} and of models of epidemic
spreading~\cite{gomez_rodriguez_inferring_2010,myers_convexity_2010,
  netrapalli_learning_2012,ma_statistical_2018,
  prasse_maximum-likelihood_2018,braunstein_alfredo_network_2019},
as well as approaches that avoid explicit modeling, such as those based
on transfer entropy~\cite{runge_escaping_2012}, Granger
causality~\cite{sun_causal_2015}, compressed
sensing~\cite{shen_reconstructing_2014, ma_efficient_2015,
han_robust_2015}, generalized
linearization~\cite{li_universal_2017}, and matching of pairwise
correlations~\cite{ching_reconstructing_2015,lai_reconstructing_2017}.

In this work, we approach the problem of network reconstruction in a
manner that is different from the aforementioned methods in two important
ways. First, we employ a nonparametric Bayesian formulation of the
problem, which yields a full posterior distribution of possible networks
that are compatible with the observed dynamical behavior. Second, we
perform network reconstruction jointly with \emph{community
detection}~\cite{fortunato_community_2016}, where at the same time as we
infer the edges of the underlying network, we also infer its modular
structure~\cite{peixoto_bayesian_2017}. As we will show, while network
reconstruction and community detection are desirable goals on their own,
joining these two tasks has a synergistic effect, whereby the detection
of communities significantly increases the accuracy of the
reconstruction, which in turn improves the discovery of the communities,
when compared to performing these tasks in isolation.

Some other approaches combine community detection with functional
observation. Berthet et al.~\cite{berthet_exact_2016} derived necessary
conditions for the exact recovery of group assignments for dense
weighted networks generated with community structure given
observed microstates of an Ising model. Hoffmann et
al.~\cite{hoffmann_community_2018} proposed a method to infer community
structure from time-series data that bypasses network reconstruction, by
employing instead a direct modeling of the dynamics given the group
assignments. However, neither of these approaches attempt to perform
network reconstruction together with community detection. Furthermore,
they are tied down to one particular inverse problem, and as we will
show, our general approach can be easily extended to an open-ended
variety of functional models.

\emph{Bayesian network reconstruction} --- We approach the network
reconstruction task similarly to the situation where the network edges
are measured directly, but via an uncertain
process~\cite{newman_network_2018-1, peixoto_reconstructing_2018}: If
$\D$ is the measurement of some process that takes place on a
network, we can define a posterior distribution for the underlying
adjacency matrix $\A$ via Bayes' rule,
\begin{equation}
  P(\A | \D) = \frac{P(\D|\A)P(\A)}{P(\D)},
\end{equation}
where $P(\D|\A)$ is an arbitrary \emph{forward} model for the dynamics
given the network, $P(\A)$ is the prior information on the network
structure, and $P(\D) = \sum_{\A}P(\D|\A)P(\A)$ is a normalization
constant comprising the total evidence for the data $\D$. We can unite
reconstruction with community detection via an, at first, seemingly minor,
but ultimately consequential modification of the above equation, where
we introduce a structured prior $P(\A|\bb)$ where $\bb$ represents the
partition of the network in communities, i.e. $\bb=\{b_i\}$, where
$b_i\in\{1,\dots,B\}$ is group membership of node $i$. This partition is
unknown, and is inferred together with the network itself, via the
joint posterior distribution
\begin{equation}\label{eq:Ab_posterior}
  P(\A,\bb | \D) = \frac{P(\D|\A)P(\A|\bb)P(\bb)}{P(\D)}.
\end{equation}
The prior $P(\A|\bb)$ is an assumed generative model for the network
structure. In our work, we will use the degree-corrected stochastic
block model (DC-SBM)~\cite{karrer_stochastic_2011}, which assumes that,
besides differences in degree, nodes belonging to the same group have
statistically equivalent connection patterns, according to the joint
probability
\begin{equation}
  P(\A|\bm\lambda,\bm\kappa,\bb) = \prod_{i<j}
  \frac{e^{-\kappa_i\kappa_j\lambda_{b_i,b_j}}(\kappa_i\kappa_j\lambda_{b_i,b_j})^{A_{ij}}}{A_{ij}!},
\end{equation}
with $\lambda_{rs}$ determining the average number of edges between
groups $r$ and $s$ and $\kappa_i$ the average degree of node $i$.
The marginal prior is obtained by integrating over all remaining parameters
weighted by their respective prior distributions,
\begin{equation}\label{sbm:marg}
  P(\A|\bb) =\int P(\A|\bm\lambda,\bm\kappa,\bb)P(\bm\kappa|\bb)P(\bm\lambda|\bb)\;\dd\bm\kappa\,\dd\bm\lambda.
\end{equation}
which can be computed exactly for standard prior choices, although it
can be modified to include hierarchical priors that have an improved
explanatory power~\cite{peixoto_nonparametric_2017} (see
Appendix~\ref{app:model} for a concise summary).

The use of the DC-SBM as a prior probability in
Eq.~\ref{eq:Ab_posterior} is motivated by its ability to inform link
prediction in networks where some fraction of edges have not been
observed or have been observed
erroneously~\cite{guimera_missing_2009,peixoto_reconstructing_2018}. The
latent conditional probabilities of edges existing between groups of
nodes is learned by the collective observation of many similar edges,
and these correlations are leveraged to extrapolate the existence of
missing or spurious ones. The same mechanism is expected to aid the
reconstruction task, where edges are not observed directly, but the
observed functional behavior yields a posterior distribution on them,
allowing the same kind of correlations to be used as an additional
source of evidence for the reconstruction, going beyond what the
dynamics alone says.

Our reconstruction approach is finalized by defining an appropriate
model for the functional behavior, determining $P(\D|\A)$. Here we will
consider two kinds of indirect data. The first comes from a SIS epidemic
spreading model~\cite{pastor-satorras_epidemic_2015}, where
$\sigma_i(t)=1$ means node $i$ is infected at time $t$, $0$
otherwise. The likelihood for this model is
\begin{equation}\label{eq:sis}
  P(\bm\sigma|\A,\btau,\gamma)=\prod_t\prod_iP(\sigma_i(t)|\bm\sigma(t-1)),
\end{equation}
where
\begin{multline}\label{eq:sis_node}
  P(\sigma_i(t)|\bm\sigma(t-1)) =\\
  f(\ee^{m_i(t-1)}, \sigma_i(t))^{1-\sigma_i(t-1)} \times f(\gamma,\sigma_i(t))^{\sigma_i(t-1)}
\end{multline}
is the transition probability for node $i$ at time $t$, with
$f(p,\sigma) = (1-p)^{\sigma}p^{1-\sigma}$, and where $m_i(t) =
\sum_jA_{ij}\ln(1-\tau_{ij})\sigma_j(t)$ is the contribution from all
neighbors of node $i$ to its infection probability at time $t$. In the
equations above the value $\tau_{ij}$ is the probability of an infection
via an existing edge $(i,j)$, and $\gamma$ is the $1\to 0$
recovery probability. With these additional parameters, the full
posterior distribution for the reconstruction becomes
\begin{equation}\label{eq:post_SIS}
  P(\A,\bb,\btau|\bm\sigma) = \frac{P(\bm\sigma|\A,\btau,\gamma)P(\A|\bb)P(\bb)P(\btau)}{P(\bm\sigma|\gamma)}.
\end{equation}
Since $\tau_{ij}\in[0,1]$ we use the uniform prior $P(\btau)=1$. Note
also that the recovery probability $\gamma$ plays no role on the
reconstruction algorithm, since its term in the likelihood does not
involve $\A$ (and hence, gets cancelled out in the denominator
$P(\bm\sigma|\gamma)=P(\gamma|\bm\sigma)P(\bm\sigma)/P(\gamma)$). This
means that the above posterior only depends on the infection events
$0\to 1$, and thus is also valid without any modifications to all
epidemic variants SI, SIR, SEIR,
etc~\cite{pastor-satorras_epidemic_2015}, since the infection events
occur with the same probability for all these models.

The second functional model we consider is the Ising model, where spin
variables on the nodes $\bm s \in \{-1,1\}^N$ are sampled according to
the joint distribution
\begin{equation}\label{eq:ising}
  P(\bm s|\A,\beta,\bm J,\bm h) = \frac{\exp\left(\beta\sum_{i<j}J_{ij}A_{ij}s_is_j + \sum_ih_is_i\right)}{Z(\A, \beta, \bm J, \bm h)},
\end{equation}
where $\beta$ is the inverse temperature, $J_{ij}$ is the coupling on
edge $(i,j)$, $h_i$ is a local field on node $i$, and $Z(\A, \beta, \bm
J, \bm h) = \sum_{\bm s}\exp(\beta\sum_{i<j}J_{ij}A_{ij}s_is_j +
\sum_ih_is_i)$ is the partition function. Note that this is not a
dynamical model, as each microstate $\bm s$ is sampled independently
according to the above distribution. Unlike the SIS model considered
before, this distribution cannot be written in closed form since $Z(\A,
\beta, \bm J, \bm h)$ cannot be computed exactly, rendering the
reconstruction problem intractable. Therefore, instead, we make use of
the pseudolikelihood approximation~\cite{besag_spatial_1974}, which is
very accurate for the purpose at hand~\cite{nguyen_inverse_2017}, where
we approximate Eq.~\ref{eq:ising} as a product of (properly normalized)
conditional probabilities for each spin variable $s_i$
\begin{equation}\label{eq:pseudo_ising}
  P(\bm s|\A,\beta,\bm J,\bm h) = \prod_i\frac{\exp(\beta s_i\sum_jJ_{ij}A_{ij}s_j + h_is_i)}{2\cosh(\beta\sum_jJ_{ij}A_{ij}s_j + h_i)}.
\end{equation}
With the above likelihood, reconstruction is performed by observing a set
of $M$ microstates $\bar{\bm s} = \{\bm s_1,\dots,\bm s_M\}$, sampled according
to $P(\bar{\bm s}|\A,\beta,\bm J,\bm h) = \prod_l P(\bm s_l|\A,\beta,\bm J,\bm h)$,
which yields the posterior distribution
\begin{multline}\label{eq:post_ising}
  P(\A,\bb, \beta,\bm J,\bm h|\bar{\bm s}) =\\
  \frac{P(\bar{\bm s}|\A,\beta,\bm J,\bm h)P(\beta)P(\bm h)P(\bm{J}|\A)P(\A|\bb)P(\bb)}{P(\bar{\bm s})}.
\end{multline}
In the above we use uniform priors $P(\bm J | \A) =
\prod_{ij}[-1/2<J_{ij}<1/2]^{A_{ij}}$, thus forcing, without loss of
generality, the values of $J_{ij}$ to lie in the shifted unit interval
$[-1/2,1/2]$. For the remaining parameters we use uniform priors, $P(\bm
h)\propto 1$ and $P(\beta)\propto 1$, for $\beta \in [-\infty, \infty]$
and $\bm h \in [-\infty, \infty]^{N}$.

For any of the above posterior distributions, we perform sampling using
Markov chain Monte Carlo (MCMC): For each proposal $\A \to \A'$, it is
accepted with the Metropolis-Hastings probability~\cite{metropolis_equation_1953,hastings_monte_1970}
\[
\min\left(1, \frac{P(\A',\bb,\bm\theta|\D)}{P(\A,\bb,\bm\theta|\D)}\frac{P(\A'\to\A)}{P(\A\to\A')} \right)
\]
and likewise for the node partition $\bb \to \bb'$, and any of the
remaining parameters $\bm\theta\to\bm\theta'$. Note that the acceptance
probability does not require the intractable normalization constant
$P(\D)$ to be computed. For both functional models considered, a whole
sweep over $E$ entries of the adjacency matrix and $N$ nodes is done in
time $O(EM + N\avg{k})$, where $M$ is the number of data samples per
node, allowing the method to be applied for large systems. We summarize
and give more details about the technical aspects of the algorithm in
Appendix~\ref{app:algo}.

\begin{figure}
  \begin{tabular}{c}
    \begin{overpic}[width=\columnwidth, trim=0 .5cm 0 0]{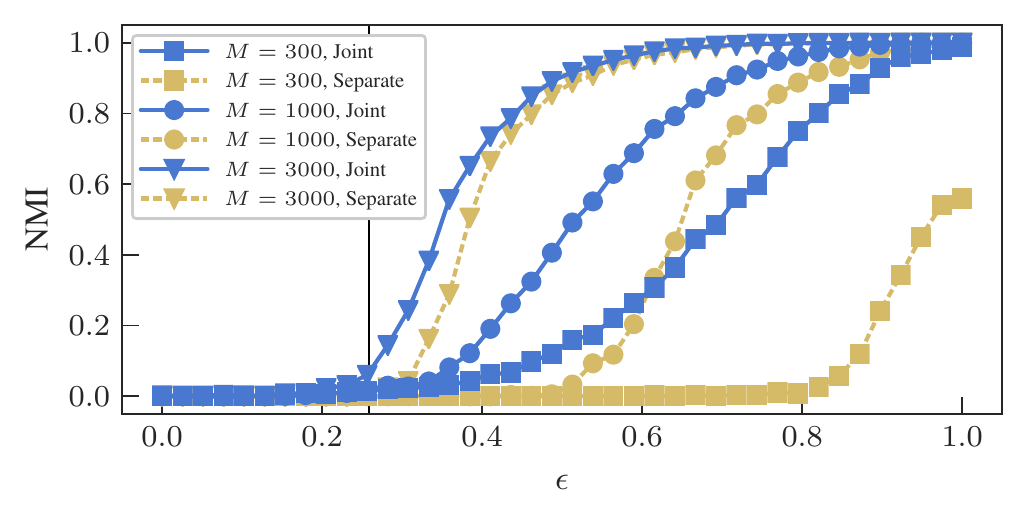}
      \put(0,0){(a)}
    \end{overpic}\\
    \begin{overpic}[width=\columnwidth, trim=0 .5cm 0 0]{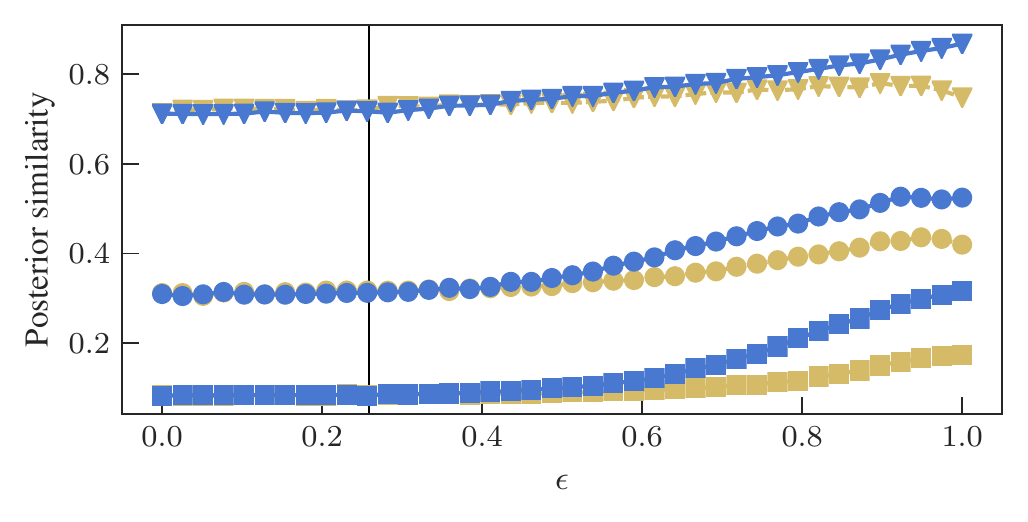}
      \put(0,0){(b)}
    \end{overpic}
  \end{tabular} \caption{\label{fig:pp}Comparison between joint and
  separate reconstruction with community detection for a PP model with
  $N=1000$, $\avg{k}=15$ and $B=10$. (a) Normalized mutual information
  (NMI) between inferred and planted node partitions, as a function of
  the model parameter $\epsilon$, for several values of the number of
  samples $M$ from the Ising model described in the text. (b) Posterior
  similarity between planted and inferred networks, for the same cases
  as in (a). The vertical line marks the detectability threshold
  $\epsilon=1/\sqrt{\avg{k}}$.}
\end{figure}

\emph{Synthetic networks} --- We begin by investigating the
reconstruction performance of networks sampled from the planted
partition model (PP), i.e. a DC-SBM with $\kappa_i=1$, $\lambda_{rs} =
\lambda_{\text{in}}\delta_{rs} + \lambda_{\text{out}}(1-\delta_{rs})$,
with $\lambda_{\text{in}}=\avg{k}(1+\epsilon(B-1))/N$ and
$\lambda_{\text{out}}=\avg{k}(1-\epsilon)/N$, where
$\epsilon=N(\lambda_{\text{in}} - \lambda_{\text{out}})/\avg{k}B$
controls the strength of the modular structure. The detectability
threshold for this model is given by $\epsilon^*=1/\sqrt{\avg{k}}$, below
which it is impossible to recover the planted community
structure~\cite{decelle_asymptotic_2011}. Given a network $\A^*$ from
this model, we sample $M$ independent Ising microstates $\bm s$
according to Eq.~\ref{eq:ising} using $J_{ij}=1$, $h_i = 0$ and
$\beta=\beta^*$ being the critical inverse temperature for the particular
network. We compare two inference approaches: In the first we sample
both the reconstructed network as well as its community structure form
the joint posterior of Eq.~\ref{eq:post_ising}. In the second approach,
we perform reconstruction and community detection separately, by first
performing reconstruction in isolation, by replacing the DC-SBM prior
$P(\A|\bb)$ by the likelihood of an Erd\H{o}s-Rényi model. We evaluate
the quality of the reconstruction via the posterior similarity
$S\in[0,1]$, defined as
\begin{equation}\label{eq:similarity}
  S(\A^*, \bm{\pi}) = 1 - \frac{\sum_{i<j}|A_{ij}^*-\pi_{ij}|}{\sum_{i<j}|A_{ij}^*+\pi_{ij}|},
\end{equation}
where $\A^*$ is the true network and $\bm\pi$ is the marginal posterior
probability for each edge, i.e. $\pi_{ij} =
\sum_{\A,\bb,\bm\theta}A_{ij}P(\A,\bb,\bm\theta|\D)$. A value $S=1$
means perfect reconstruction. We then perform community
detection \emph{a posteriori} by obtaining the maximum marginal point estimate
\begin{equation}
  \hat{A}_{ij} =
  \begin{cases}
    1 & \text{if } \pi_{ij} > 1/2, \\
    0 & \text{if } \pi_{ij} < 1/2.
  \end{cases}
\end{equation}
and then sampling from the posterior $P(\bb|\hat\A)$. Fig.~\ref{fig:pp}
contains the comparison between both approaches for networks sampled
from the PP model, which shows how sampling from the joint posterior
improves both the reconstruction as well as community detection. For the
latter, the joint inference allows the detection all the way down to the
detectability threshold, for the examples considered, which, otherwise, is
not possible with the separate method.

\begin{figure}
  \begin{tabular}{cc}
     \begin{overpic}[width=.5\columnwidth, trim=0 .5cm 0 0]{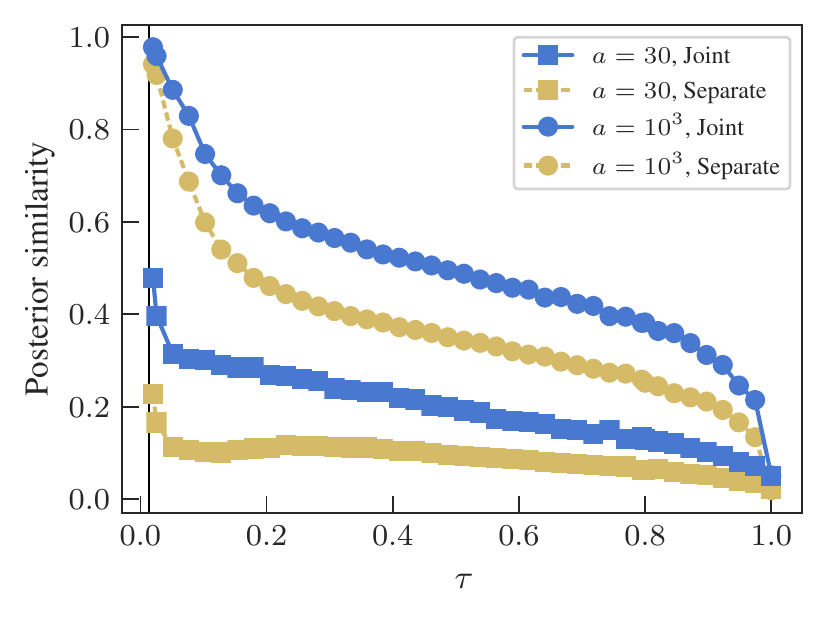}
      \put(0,0){(a)}
      \end{overpic}&
     \begin{overpic}[width=.5\columnwidth, trim=0 .5cm 0 0]{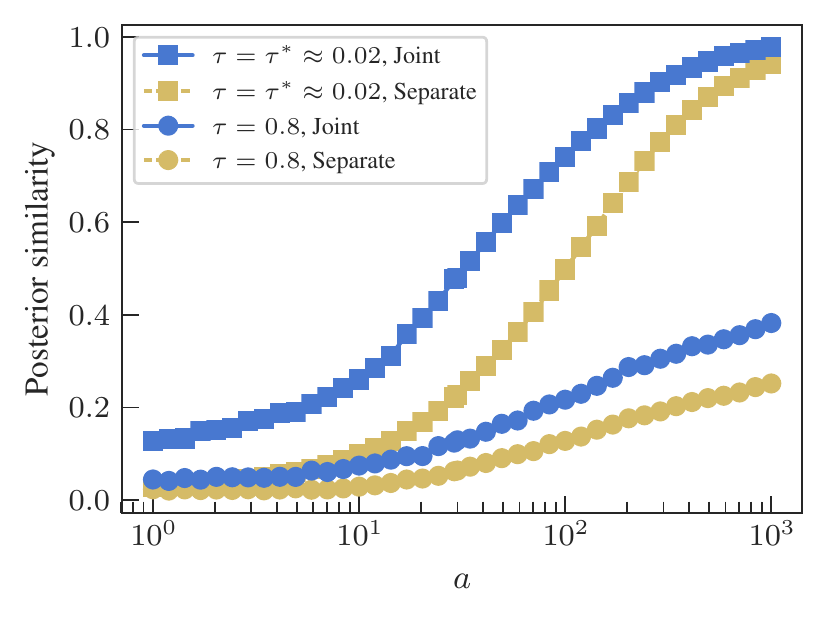}
      \put(0,0){(b)}
      \end{overpic}\\
     \begin{overpic}[width=.5\columnwidth, trim=0 .5cm 0 0]{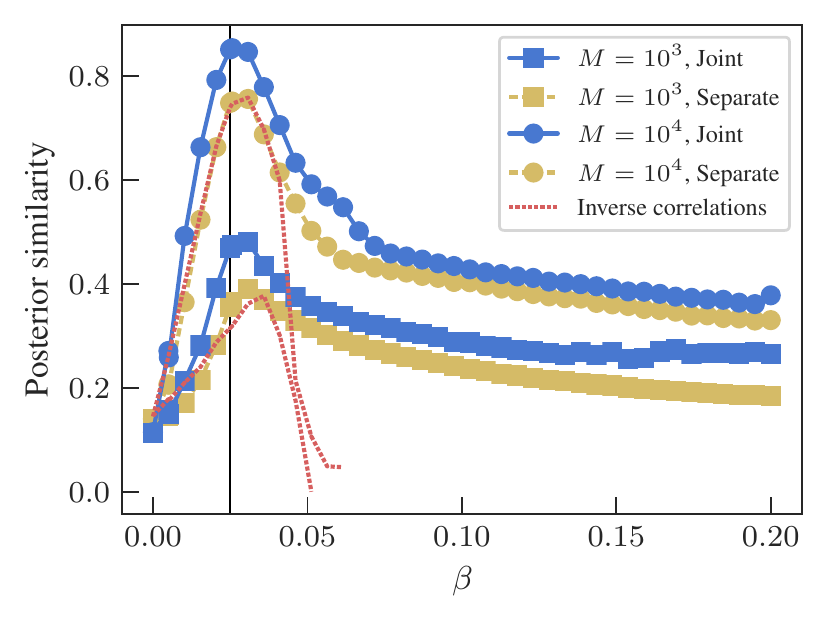}
      \put(0,0){(c)}
      \end{overpic}&
     \begin{overpic}[width=.5\columnwidth, trim=0 .5cm 0 0]{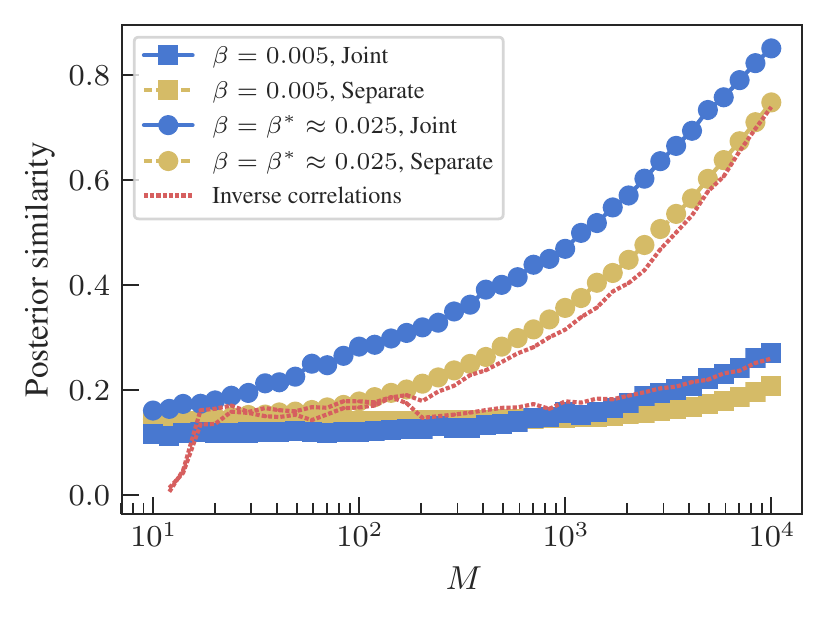}
      \put(0,0){(d)}
      \end{overpic}
  \end{tabular} \caption{\label{fig:similarity} Reconstruction results
  for simulated dynamics on empirical networks, comparing separate and
  joint reconstruction with community detection. (a) and (b) correspond
  to a SIS dynamics on global airport data, using $\tau_{ij}=\tau$,
  $\gamma=1$, for different values of the infection probability $\tau$
  and node activity $a$ (defined as the number of infection events per
  node), and (c) and (d) the Ising model on a food web, using $J_{ij}=1$
  and $h_i=0$. The dashed red line corresponds to the inverse
  correlation method for the Ising model. The solid vertical line marks
  the critical value for each model.}
\end{figure}

\emph{Real networks with synthetic dynamics} --- Now, we investigate the
reconstruction of networks not generated by the DC-SBM. We take two
empirical networks, the worldwide network of $N=3\,286$
airports~\footnote{Retrieved from \url{openflights.org}.} with
$E=39\,430$ edges, and a food web from Little Rock
Lake~\cite{martinez_artifacts_1991}, containing $N=183$ nodes and
$E=2\,434$ edges, and we sample from the SIS (mimicking the spread of a
pandemic) and Ising model (representing simplified inter-species
interactions) on them, respectively, and evaluate the reconstruction
obtained via the joint and separate inference with community detection,
with results shown in Fig.~\ref{fig:similarity}. As is also the case for
synthetic networks, the reconstruction quality is significantly improved
by performing joint community detection~\footnote{Note that in this case
our method also exploits the heterogeneous degrees in the network via
the DC-SBM, which can in principle also aid the reconstruction, in
addition to the community structure itself.}. The quality of the
reconstruction peaks at the critical threshold for each model, at which
the sensitivity to perturbations is the largest. As the number of
observed samples increases, so does the quality of the reconstruction,
and the relative advantage of the joint reconstruction diminishes, as
the data eventually ``washes out'' the contribution from the prior. For
the Ising model, we compare the results of our method with the
mean-field inverse correlations method~\cite{nguyen_inverse_2017},
i.e. $\beta A_{ij}J_{ij} = [\bm
  C^{-1}]_{ij}$, where $C_{ij} = \avg{\sigma_i\sigma_j} -
\avg{\sigma_i}\avg{\sigma_j}$ is the connected correlation matrix. As
seen in Fig.~\ref{fig:similarity}, this simpler reconstruction method
can be just as accurate as our separate reconstruction approach, but
only close to the critical point. For higher inverse temperatures the
reconstruction deteriorates rapidly, and breaks down completely as the
system becomes locally magnetized, with whole rows and columns of the
matrix $\bm C$ being equal to zero, causing it to be
singular~\footnote{Refinements of this approach including
  TAP and BP corrections~\cite{nguyen_inverse_2017} yield the same
  performance for this example.}. In such situations this kind of
approach requires explicit regularization
techniques~\cite{decelle_pseudolikelihood_2014}, which become
unnecessary with our Bayesian method. The joint inference with community
structure improves the reconstruction even further, beyond what is
obtainable with typical inverse Ising methods, since it incorporates a
different source of evidence.

\begin{figure}
  \begin{tabular}{cc}
    \begin{overpic}[width=.45\columnwidth,trim=.2cm 1.5cm .2cm 2.4cm, clip]{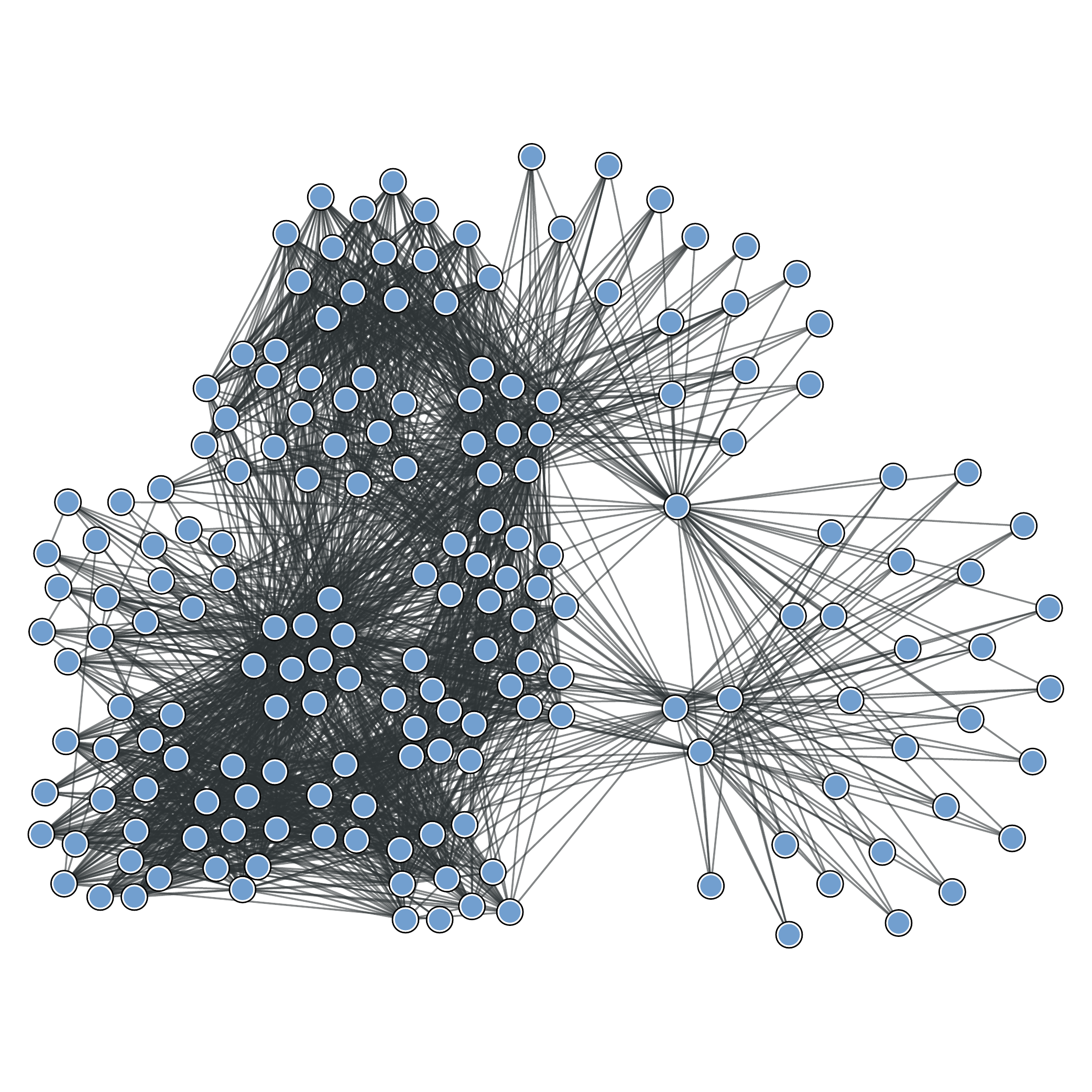}
      \put(0,0){(a)}
      \end{overpic}&
    \begin{overpic}[width=.45\columnwidth,trim=.2cm 1.5cm .2cm 2.4cm, clip]{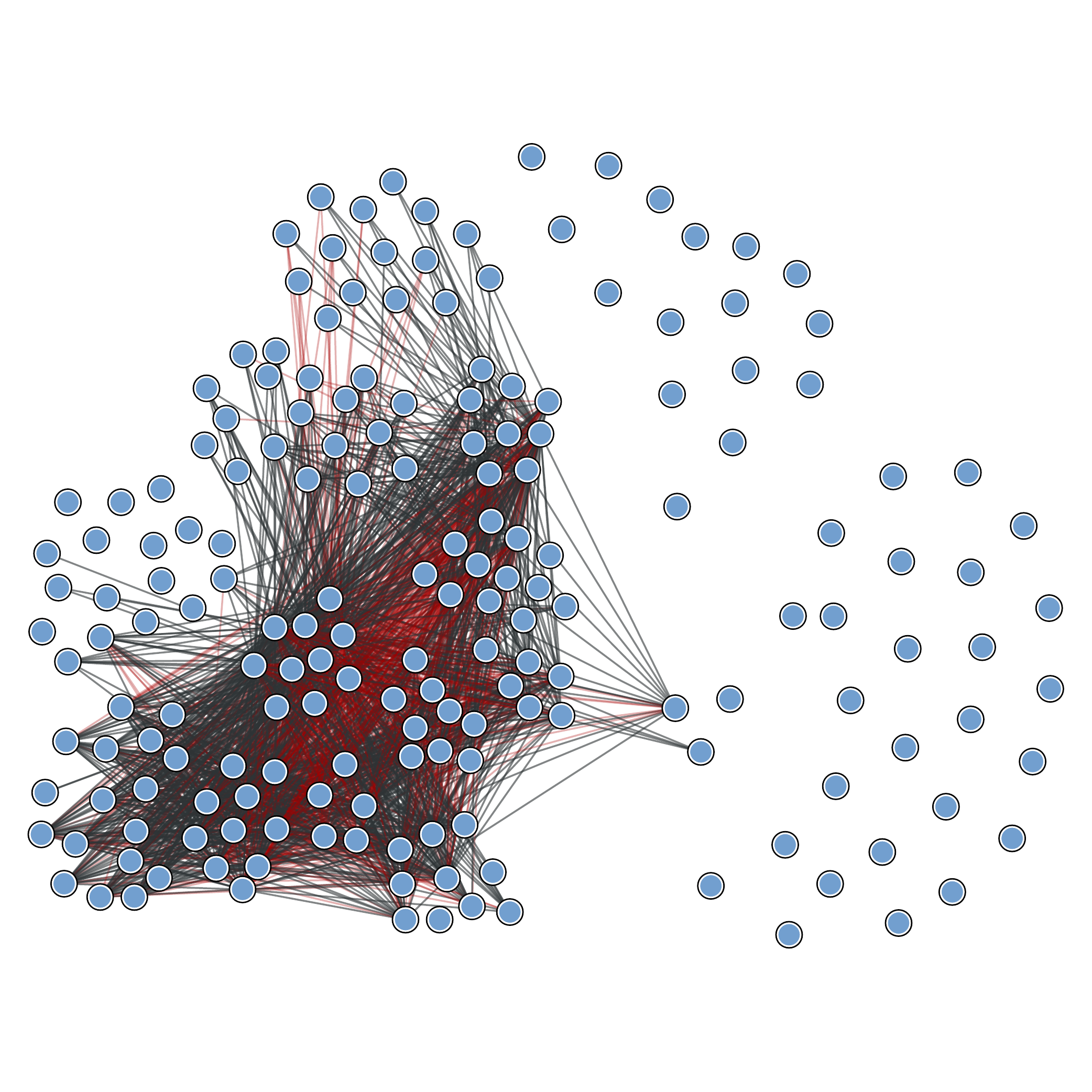}
      \put(0,0){(b)\hspace{2.5em} S = 0.66}
      \end{overpic}\\
    \begin{overpic}[width=.45\columnwidth,trim=.2cm 1.5cm .2cm 2.4cm, clip]{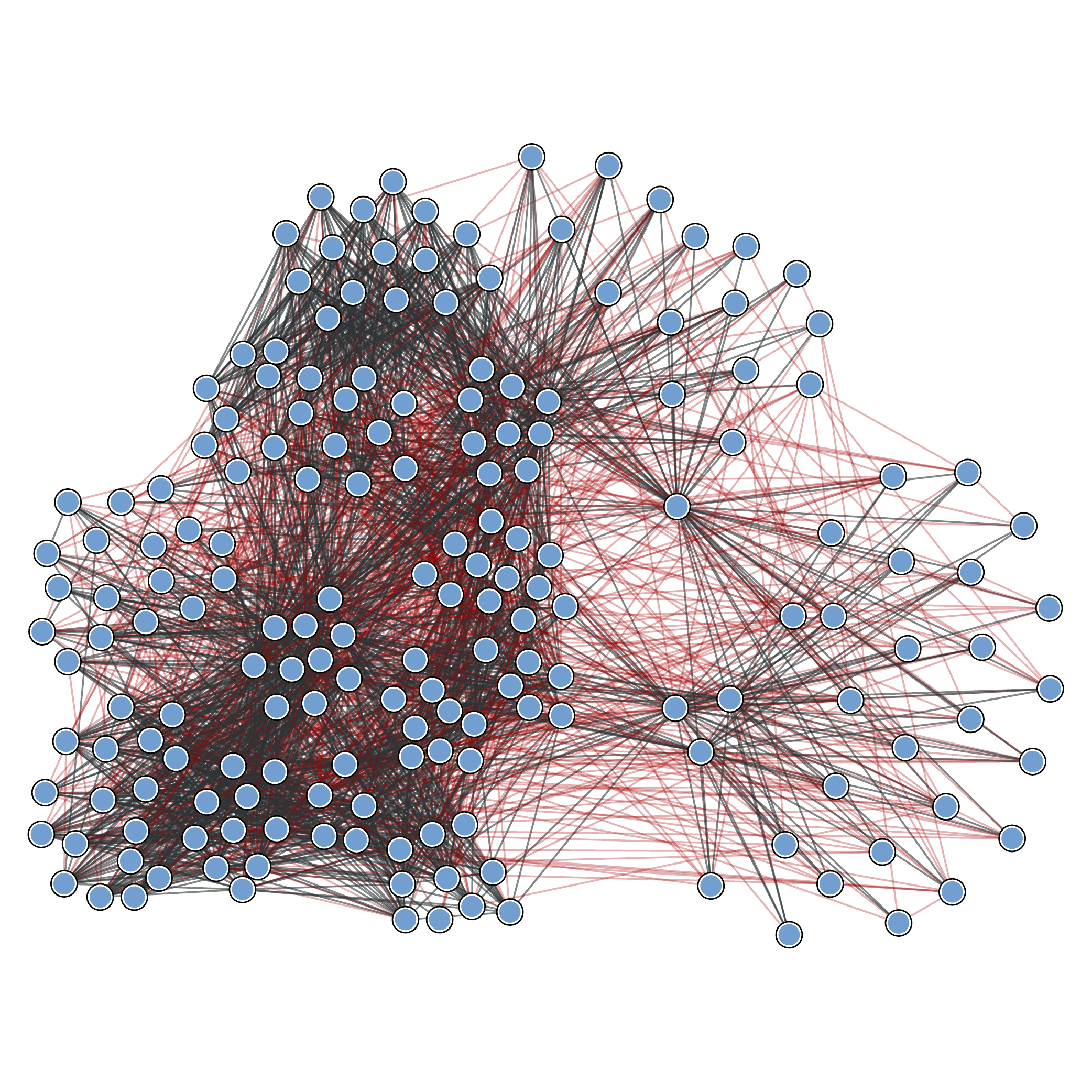}
      \put(0,0){(c)\hspace{2.5em} S = 0.74}
      \end{overpic}&
    \begin{overpic}[width=.45\columnwidth,trim=.2cm 1.5cm .2cm 2.4cm, clip]{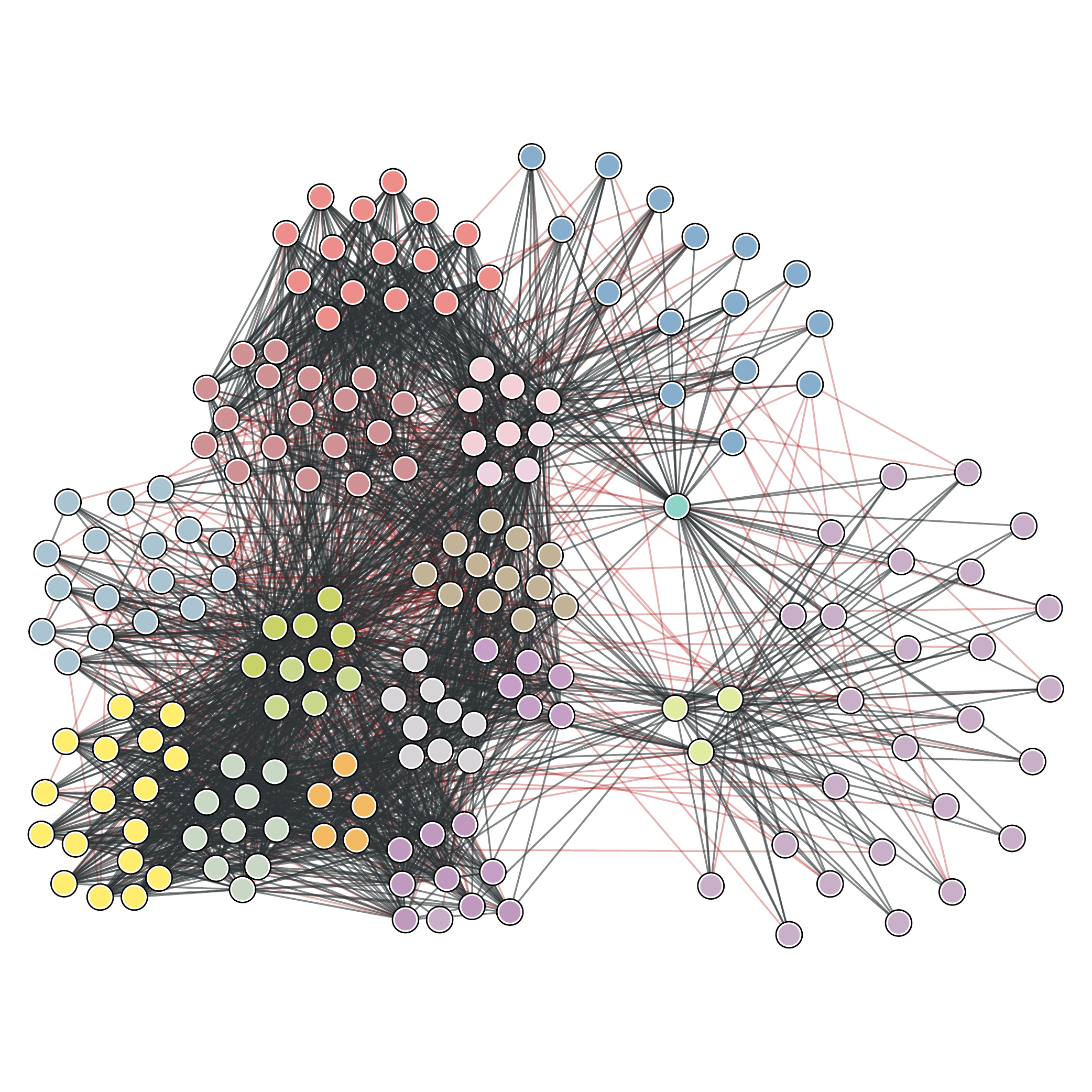}
      \put(0,0){(d)\hspace{2.5em} S = 0.88}
      \end{overpic}
  \end{tabular} \caption{\label{fig:thres}Reconstruction of a food web
  network~\cite{martinez_artifacts_1991} from $M=10^4$ samples of an
  Ising model at critical temperature. Edges marked in red are erroneous
  in the reconstruction. (a) Original network, (b) Optimal correlation
  thresholding, (c) Inverse correlations, (d) Joint reconstruction with
  community detection. The legends show the values of the posterior
  similarity (Eq.~\ref{eq:similarity}).}
\end{figure}

In Fig.~\ref{fig:thres} we show a comparison of the reconstruction of
the food web network from a simulated Ising model, using different
approaches. Optimal thresholding corresponds to the naive approach of
imputing the existence of an edge to the connected correlation between
two nodes exceeding a threshold $c^*$, i.e. $\pi_{ij}=\{1 \text{ if }
C_{ij}>c^*, \; 0 \text{ otherwise}\}$. The value of $c^*$ was chosen to maximize
the posterior similarity, which represents the best possible
reconstruction achievable with this method. Nevertheless, the network
thus obtained is severely distorted. The inverse correlation method
comes much closer to the true network, but is superseded by the joint
inference with community detection.

\begin{figure}
  \begin{tabular}{c}
    \begin{overpic}[width=\columnwidth, trim=0 0 .5cm 0]{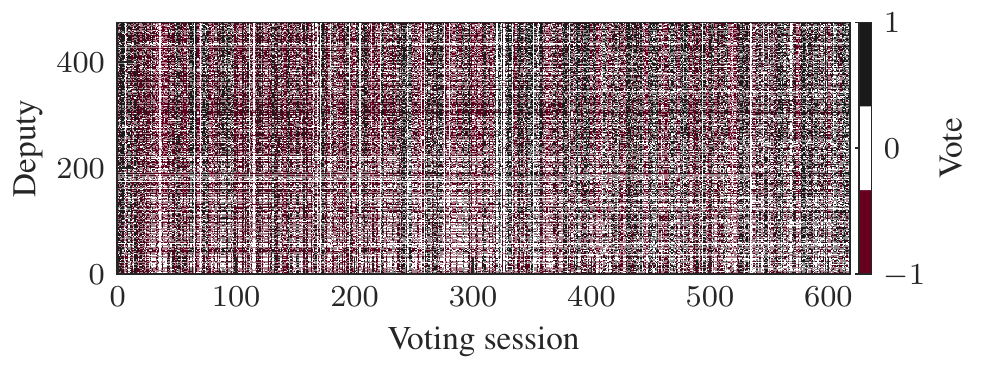}
      \put(0,35){(a)}
    \end{overpic}\\
    \begin{overpic}[width=\columnwidth, trim=0 0 0 -1cm]{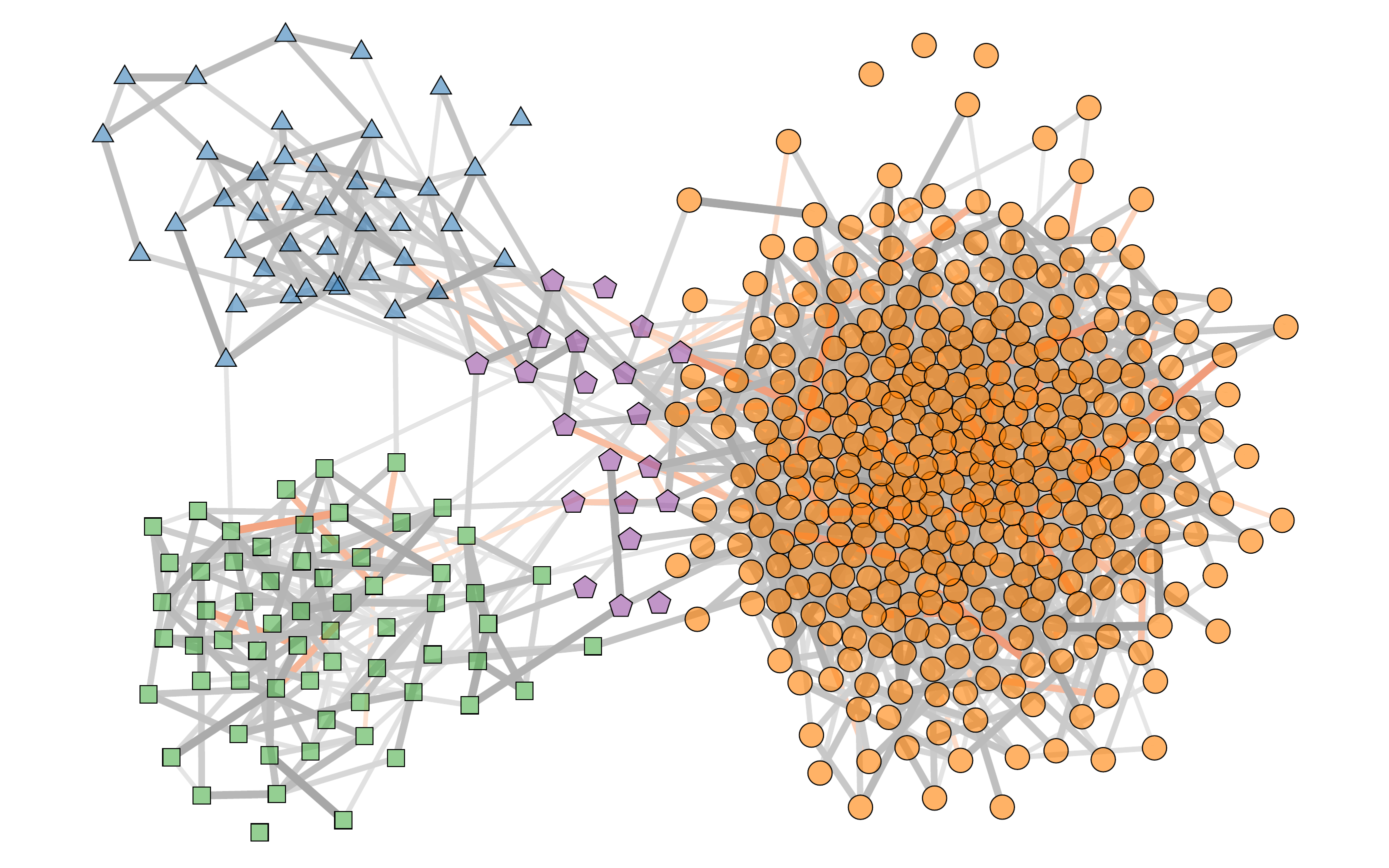}
      \put(0,65){(b)}
      \put(13, 65){\larger[2] Opposition}
      \put(58, 65){\larger[2] Government}
      \put(34, 57){\smaller[2] DEM}
      \put(31, 05){\smaller[2] PSDB}
      \put(42, 12){\smaller[2] \begin{minipage}{1cm}DEM, PMDB\end{minipage}}
      \put(78.5, 52.5){\begin{minipage}{2cm}\smaller[2] PT, PMDB, PDT, PTB, PCdoB, PP, PR, PV\end{minipage}}
    \end{overpic}
  \end{tabular} \caption{\label{fig:congress}Reconstruction of the
  interactions between members of the lower house of the Brazilian
  congress from the voting patterns of the 2007-2011 session, according
  to the Ising model. 
  The node colors indicate the inferred groups. The edge thickness shows
  the posterior probability for each edge, and the color the magnitude
  of the coupling $J_{ij}$. The labels show the most frequent party
  membership for each group.}
\end{figure}

\emph{Empirical dynamics} --- We turn to the reconstruction from
observed empirical dynamics with unknown underlying interactions. The
first example is the sequence of $M=619$ votes of $N=575$ deputies in
the 2007 to 2011 session of the lower chamber of the Brazilian
congress. Each deputy voted Yes, No, or abstained for each legislation,
which we represent as $\{1,-1,0\}$, respectively. Since the temporal
ordering of the voting sessions is likely to be of secondary importance
to the voting outcomes, we assume the votes are sampled from an Ising
model (the addition of zero-valued spins changes
Eq.~\ref{eq:pseudo_ising} only slightly by replacing $2\cosh(x)\to
1+2\cosh(x)$). Fig~\ref{fig:congress} shows the result of the
reconstruction, where the division of the nodes uncovers a cohesive
government and a split opposition, as well as a marginal center group,
which correlates very well with the known party memberships and can be
use to predict unseen voting behavior (see
Appendix~\ref{app:empirical}). In Fig~\ref{fig:twitter} we show the
result of the reconstruction of the directed network of influence
between $N=1\,833$ twitter users from $58\,224$
re-tweets~\cite{hodas_simple_2014} using a SI epidemic model (the act of
``re-tweeting'' is modelled as an infection event, using
Eqs.~\ref{eq:sis} and~\ref{eq:sis_node} with $\gamma=0$) and the nested
DC-SBM. The reconstruction uncovers isolated groups with varying
propensities to re-tweet, as well as groups that tend to be influence a
large fraction of users. By inspecting the geo-location metadata on the
users, we see that the inferred groups amount to a large extent do
different countries, although clear sub-divisions indicate that this is
not the only factor governing the influence among users (see
Appendix~\ref{app:metadata}).

\begin{figure}

  \hspace{-1cm}\begin{overpic}[width=.8\columnwidth]{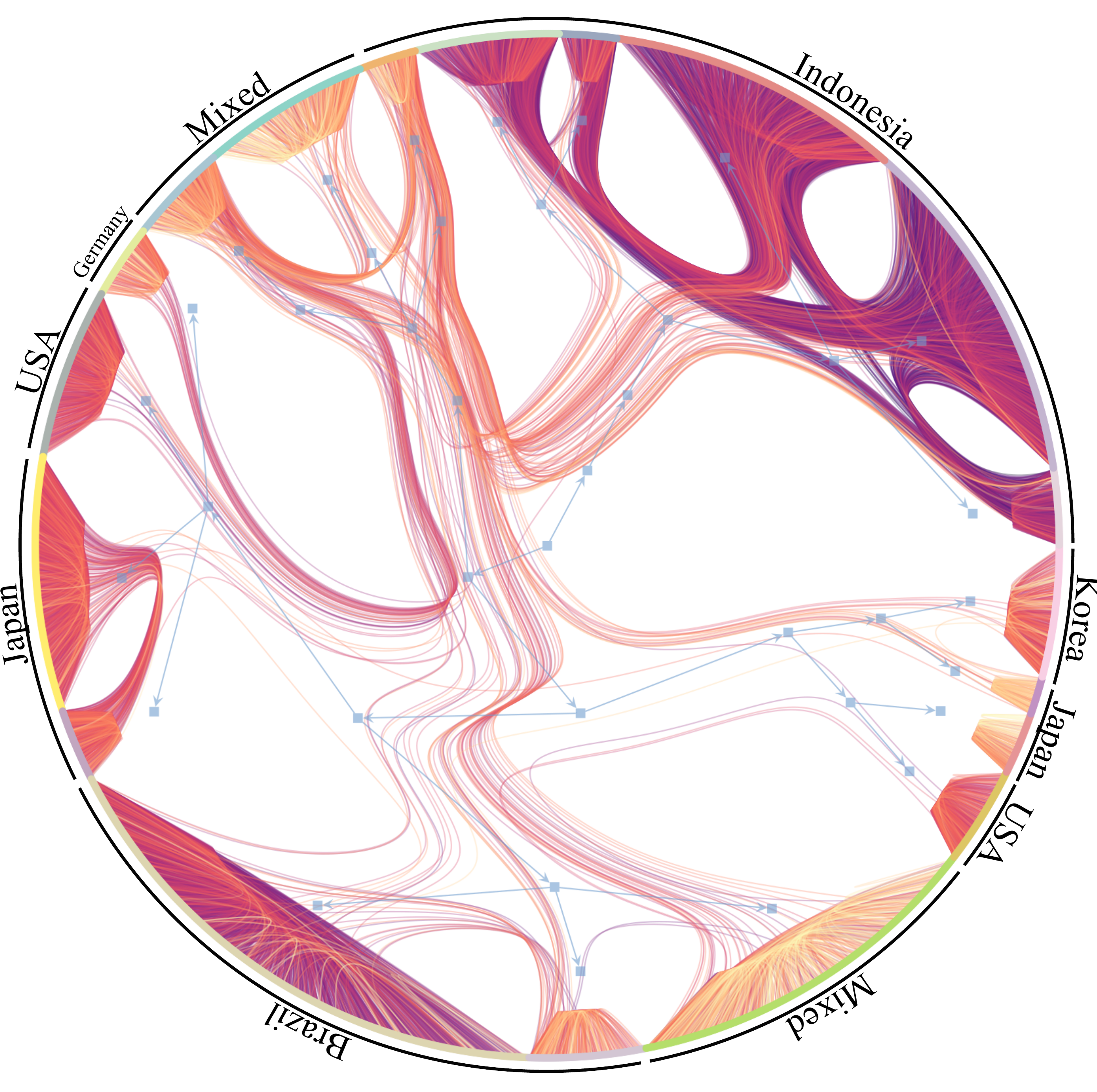}
    \put(105,0){\rotatebox{90}{\includegraphics[width=.8\columnwidth]{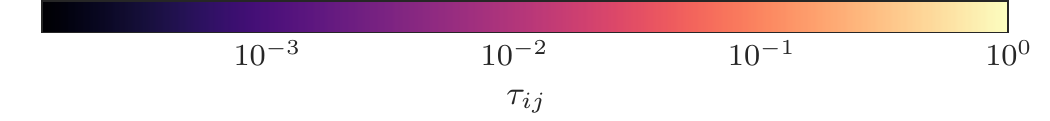}}}
  \end{overpic}
  \caption{\label{fig:twitter} Reconstruction of the directed network of
  influence between
    $N=1\,833$ twitter users from $58\,224$ re-tweets, using a SI infection
  model. The hierarchical division represents the inferred fit of the
  nested DC-SBM (see
  Refs.~\cite{peixoto_hierarchical_2014,holten_hierarchical_2006} for
  details on the layout algorithm), and the edge colors indicate the
  infection probabilities $\tau_{ij}$ as shown in the legend. The text
  labels show the dominating country membership for the users in each
  group. }
\end{figure}

\emph{Conclusion} --- We have presented a scalable Bayesian method to
reconstruct networks from functional observations that uses the SBM as a
structured prior, and, hence, performs community detection together with
reconstruction. The method is nonparametric, and, hence, requires no prior
stipulation of aspects of the network and size of the model, such as
number of groups. By leveraging inferred correlations between edges, the
SBM includes an additional source of evidence, and, thereby, improves the
reconstruction accuracy, which in turn also increases the accuracy of
the inferred communities. The overall approach is general, requiring
only appropriate functional model specifications, and can be coupled
with an open ended variety of such models, other than those considered
here.

\bibliography{bib,extra}

\begin{thebibliography}{52}%
\makeatletter
\providecommand \@ifxundefined [1]{%
 \@ifx{#1\undefined}
}%
\providecommand \@ifnum [1]{%
 \ifnum #1\expandafter \@firstoftwo
 \else \expandafter \@secondoftwo
 \fi
}%
\providecommand \@ifx [1]{%
 \ifx #1\expandafter \@firstoftwo
 \else \expandafter \@secondoftwo
 \fi
}%
\providecommand \natexlab [1]{#1}%
\providecommand \enquote  [1]{``#1''}%
\providecommand \bibnamefont  [1]{#1}%
\providecommand \bibfnamefont [1]{#1}%
\providecommand \citenamefont [1]{#1}%
\providecommand \href@noop [0]{\@secondoftwo}%
\providecommand \href [0]{\begingroup \@sanitize@url \@href}%
\providecommand \@href[1]{\@@startlink{#1}\@@href}%
\providecommand \@@href[1]{\endgroup#1\@@endlink}%
\providecommand \@sanitize@url [0]{\catcode `\\12\catcode `\$12\catcode
  `\&12\catcode `\#12\catcode `\^12\catcode `\_12\catcode `\%12\relax}%
\providecommand \@@startlink[1]{}%
\providecommand \@@endlink[0]{}%
\providecommand \url  [0]{\begingroup\@sanitize@url \@url }%
\providecommand \@url [1]{\endgroup\@href {#1}{\urlprefix }}%
\providecommand \urlprefix  [0]{URL }%
\providecommand \Eprint [0]{\href }%
\providecommand \doibase [0]{http://dx.doi.org/}%
\providecommand \selectlanguage [0]{\@gobble}%
\providecommand \bibinfo  [0]{\@secondoftwo}%
\providecommand \bibfield  [0]{\@secondoftwo}%
\providecommand \translation [1]{[#1]}%
\providecommand \BibitemOpen [0]{}%
\providecommand \bibitemStop [0]{}%
\providecommand \bibitemNoStop [0]{.\EOS\space}%
\providecommand \EOS [0]{\spacefactor3000\relax}%
\providecommand \BibitemShut  [1]{\csname bibitem#1\endcsname}%
\let\auto@bib@innerbib\@empty
\bibitem [{\citenamefont {Wang}\ \emph {et~al.}(2006)\citenamefont {Wang},
  \citenamefont {Joshi}, \citenamefont {Zhang}, \citenamefont {Xu},\ and\
  \citenamefont {Chen}}]{wang_inferring_2006}%
  \BibitemOpen
  \bibfield  {author} {\bibinfo {author} {\bibfnamefont {Yong}\ \bibnamefont
  {Wang}}, \bibinfo {author} {\bibfnamefont {Trupti}\ \bibnamefont {Joshi}},
  \bibinfo {author} {\bibfnamefont {Xiang-Sun}\ \bibnamefont {Zhang}}, \bibinfo
  {author} {\bibfnamefont {Dong}\ \bibnamefont {Xu}}, \ and\ \bibinfo {author}
  {\bibfnamefont {Luonan}\ \bibnamefont {Chen}},\ }\bibfield  {title}
  {{\selectlanguage {english}\enquote {\bibinfo {title} {Inferring gene
  regulatory networks from multiple microarray datasets},}\ }}\href {\doibase
  10.1093/bioinformatics/btl396} {\bibfield  {journal} {\bibinfo  {journal}
  {Bioinformatics}\ }\textbf {\bibinfo {volume} {22}},\ \bibinfo {pages}
  {2413--2420} (\bibinfo {year} {2006})}\BibitemShut {NoStop}%
\bibitem [{\citenamefont {Breakspear}(2017)}]{breakspear_dynamic_2017}%
  \BibitemOpen
  \bibfield  {author} {\bibinfo {author} {\bibfnamefont {Michael}\ \bibnamefont
  {Breakspear}},\ }\bibfield  {title} {{\selectlanguage {english}\enquote
  {\bibinfo {title} {Dynamic models of large-scale brain activity},}\ }}\href
  {\doibase 10.1038/nn.4497} {\bibfield  {journal} {\bibinfo  {journal} {Nature
  Neuroscience}\ }\textbf {\bibinfo {volume} {20}},\ \bibinfo {pages}
  {340--352} (\bibinfo {year} {2017})}\BibitemShut {NoStop}%
\bibitem [{\citenamefont {Keeling}\ and\ \citenamefont
  {Rohani}(2002)}]{keeling_estimating_2002}%
  \BibitemOpen
  \bibfield  {author} {\bibinfo {author} {\bibfnamefont {Matt~J.}\ \bibnamefont
  {Keeling}}\ and\ \bibinfo {author} {\bibfnamefont {Pejman}\ \bibnamefont
  {Rohani}},\ }\bibfield  {title} {{\selectlanguage {english}\enquote {\bibinfo
  {title} {Estimating spatial coupling in epidemiological systems: a
  mechanistic approach},}\ }}\href {\doibase 10.1046/j.1461-0248.2002.00268.x}
  {\bibfield  {journal} {\bibinfo  {journal} {Ecology Letters}\ }\textbf
  {\bibinfo {volume} {5}},\ \bibinfo {pages} {20--29} (\bibinfo {year}
  {2002})}\BibitemShut {NoStop}%
\bibitem [{\citenamefont {Musmeci}\ \emph {et~al.}(2013)\citenamefont
  {Musmeci}, \citenamefont {Battiston}, \citenamefont {Caldarelli},
  \citenamefont {Puliga},\ and\ \citenamefont
  {Gabrielli}}]{musmeci_bootstrapping_2013}%
  \BibitemOpen
  \bibfield  {author} {\bibinfo {author} {\bibfnamefont {Nicolò}\ \bibnamefont
  {Musmeci}}, \bibinfo {author} {\bibfnamefont {Stefano}\ \bibnamefont
  {Battiston}}, \bibinfo {author} {\bibfnamefont {Guido}\ \bibnamefont
  {Caldarelli}}, \bibinfo {author} {\bibfnamefont {Michelangelo}\ \bibnamefont
  {Puliga}}, \ and\ \bibinfo {author} {\bibfnamefont {Andrea}\ \bibnamefont
  {Gabrielli}},\ }\bibfield  {title} {{\selectlanguage {english}\enquote
  {\bibinfo {title} {Bootstrapping {Topological} {Properties} and {Systemic}
  {Risk} of {Complex} {Networks} {Using} the {Fitness} {Model}},}\ }}\href
  {\doibase 10.1007/s10955-013-0720-1} {\bibfield  {journal} {\bibinfo
  {journal} {Journal of Statistical Physics}\ }\textbf {\bibinfo {volume}
  {151}},\ \bibinfo {pages} {720--734} (\bibinfo {year} {2013})}\BibitemShut
  {NoStop}%
\bibitem [{\citenamefont {Bakshy}\ \emph {et~al.}(2012)\citenamefont {Bakshy},
  \citenamefont {Rosenn}, \citenamefont {Marlow},\ and\ \citenamefont
  {Adamic}}]{bakshy_role_2012}%
  \BibitemOpen
  \bibfield  {author} {\bibinfo {author} {\bibfnamefont {Eytan}\ \bibnamefont
  {Bakshy}}, \bibinfo {author} {\bibfnamefont {Itamar}\ \bibnamefont {Rosenn}},
  \bibinfo {author} {\bibfnamefont {Cameron}\ \bibnamefont {Marlow}}, \ and\
  \bibinfo {author} {\bibfnamefont {Lada}\ \bibnamefont {Adamic}},\ }\bibfield
  {title} {\enquote {\bibinfo {title} {The {Role} of {Social} {Networks} in
  {Information} {Diffusion}},}\ }in\ \href {\doibase 10.1145/2187836.2187907}
  {\emph {\bibinfo {booktitle} {Proceedings of the 21st {International}
  {Conference} on {World} {Wide} {Web}}}},\ \bibinfo {series and number} {{WWW}
  '12}\ (\bibinfo  {publisher} {ACM},\ \bibinfo {address} {New York, NY, USA},\
  \bibinfo {year} {2012})\ pp.\ \bibinfo {pages} {519--528},\ \bibinfo {note}
  {event-place: Lyon, France}\BibitemShut {NoStop}%
\bibitem [{\citenamefont {Kramer}\ \emph {et~al.}(2009)\citenamefont {Kramer},
  \citenamefont {Eden}, \citenamefont {Cash},\ and\ \citenamefont
  {Kolaczyk}}]{kramer_network_2009}%
  \BibitemOpen
  \bibfield  {author} {\bibinfo {author} {\bibfnamefont {Mark~A.}\ \bibnamefont
  {Kramer}}, \bibinfo {author} {\bibfnamefont {Uri~T.}\ \bibnamefont {Eden}},
  \bibinfo {author} {\bibfnamefont {Sydney~S.}\ \bibnamefont {Cash}}, \ and\
  \bibinfo {author} {\bibfnamefont {Eric~D.}\ \bibnamefont {Kolaczyk}},\
  }\bibfield  {title} {\enquote {\bibinfo {title} {Network inference with
  confidence from multivariate time series},}\ }\href {\doibase
  10.1103/PhysRevE.79.061916} {\bibfield  {journal} {\bibinfo  {journal}
  {Physical Review E}\ }\textbf {\bibinfo {volume} {79}},\ \bibinfo {pages}
  {061916} (\bibinfo {year} {2009})}\BibitemShut {NoStop}%
\bibitem [{\citenamefont {Timme}(2007)}]{timme_revealing_2007}%
  \BibitemOpen
  \bibfield  {author} {\bibinfo {author} {\bibfnamefont {Marc}\ \bibnamefont
  {Timme}},\ }\bibfield  {title} {\enquote {\bibinfo {title} {Revealing
  {Network} {Connectivity} from {Response} {Dynamics}},}\ }\href {\doibase
  10.1103/PhysRevLett.98.224101} {\bibfield  {journal} {\bibinfo  {journal}
  {Physical Review Letters}\ }\textbf {\bibinfo {volume} {98}},\ \bibinfo
  {pages} {224101} (\bibinfo {year} {2007})}\BibitemShut {NoStop}%
\bibitem [{\citenamefont {Shandilya}\ and\ \citenamefont
  {Timme}(2011)}]{shandilya_inferring_2011}%
  \BibitemOpen
  \bibfield  {author} {\bibinfo {author} {\bibfnamefont {Srinivas~Gorur}\
  \bibnamefont {Shandilya}}\ and\ \bibinfo {author} {\bibfnamefont {Marc}\
  \bibnamefont {Timme}},\ }\bibfield  {title} {{\selectlanguage
  {english}\enquote {\bibinfo {title} {Inferring network topology from complex
  dynamics},}\ }}\href {\doibase 10.1088/1367-2630/13/1/013004} {\bibfield
  {journal} {\bibinfo  {journal} {New Journal of Physics}\ }\textbf {\bibinfo
  {volume} {13}},\ \bibinfo {pages} {013004} (\bibinfo {year}
  {2011})}\BibitemShut {NoStop}%
\bibitem [{\citenamefont {Nitzan}\ \emph {et~al.}(2017)\citenamefont {Nitzan},
  \citenamefont {Casadiego},\ and\ \citenamefont
  {Timme}}]{nitzan_revealing_2017}%
  \BibitemOpen
  \bibfield  {author} {\bibinfo {author} {\bibfnamefont {Mor}\ \bibnamefont
  {Nitzan}}, \bibinfo {author} {\bibfnamefont {Jose}\ \bibnamefont
  {Casadiego}}, \ and\ \bibinfo {author} {\bibfnamefont {Marc}\ \bibnamefont
  {Timme}},\ }\bibfield  {title} {{\selectlanguage {english}\enquote {\bibinfo
  {title} {Revealing physical interaction networks from statistics of
  collective dynamics},}\ }}\href {\doibase 10.1126/sciadv.1600396} {\bibfield
  {journal} {\bibinfo  {journal} {Science Advances}\ }\textbf {\bibinfo
  {volume} {3}},\ \bibinfo {pages} {e1600396} (\bibinfo {year}
  {2017})}\BibitemShut {NoStop}%
\bibitem [{\citenamefont {Abbeel}\ \emph {et~al.}(2006)\citenamefont {Abbeel},
  \citenamefont {Koller},\ and\ \citenamefont {Ng}}]{abbeel_learning_2006}%
  \BibitemOpen
  \bibfield  {author} {\bibinfo {author} {\bibfnamefont {Pieter}\ \bibnamefont
  {Abbeel}}, \bibinfo {author} {\bibfnamefont {Daphne}\ \bibnamefont {Koller}},
  \ and\ \bibinfo {author} {\bibfnamefont {Andrew~Y.}\ \bibnamefont {Ng}},\
  }\bibfield  {title} {\enquote {\bibinfo {title} {Learning {Factor} {Graphs}
  in {Polynomial} {Time} and {Sample} {Complexity}},}\ }\href
  {http://www.jmlr.org/papers/v7/abbeel06a.html} {\bibfield  {journal}
  {\bibinfo  {journal} {Journal of Machine Learning Research}\ }\textbf
  {\bibinfo {volume} {7}},\ \bibinfo {pages} {1743--1788} (\bibinfo {year}
  {2006})}\BibitemShut {NoStop}%
\bibitem [{\citenamefont {Bresler}\ \emph {et~al.}(2008)\citenamefont
  {Bresler}, \citenamefont {Mossel},\ and\ \citenamefont
  {Sly}}]{bresler_reconstruction_2008}%
  \BibitemOpen
  \bibfield  {author} {\bibinfo {author} {\bibfnamefont {Guy}\ \bibnamefont
  {Bresler}}, \bibinfo {author} {\bibfnamefont {Elchanan}\ \bibnamefont
  {Mossel}}, \ and\ \bibinfo {author} {\bibfnamefont {Allan}\ \bibnamefont
  {Sly}},\ }\bibfield  {title} {{\selectlanguage {english}\enquote {\bibinfo
  {title} {Reconstruction of {Markov} {Random} {Fields} from {Samples}: {Some}
  {Observations} and {Algorithms}},}\ }}in\ \href {\doibase
  10.1007/978-3-540-85363-3_28} {{\selectlanguage {english}\emph {\bibinfo
  {booktitle} {Approximation, {Randomization} and {Combinatorial}
  {Optimization}. {Algorithms} and {Techniques}}}}},\ \bibinfo {series and
  number} {Lecture {Notes} in {Computer} {Science}}\ (\bibinfo  {publisher}
  {Springer, Berlin, Heidelberg},\ \bibinfo {year} {2008})\ pp.\ \bibinfo
  {pages} {343--356}\BibitemShut {NoStop}%
\bibitem [{\citenamefont {Montanari}\ and\ \citenamefont
  {Pereira}(2009)}]{montanari_which_2009}%
  \BibitemOpen
  \bibfield  {author} {\bibinfo {author} {\bibfnamefont {Andrea}\ \bibnamefont
  {Montanari}}\ and\ \bibinfo {author} {\bibfnamefont {Jose~A.}\ \bibnamefont
  {Pereira}},\ }\bibfield  {title} {\enquote {\bibinfo {title} {Which graphical
  models are difficult to learn?}}\ }in\ \href
  {http://papers.nips.cc/paper/3819-which-graphical-models-are-difficult-to-learn.pdf}
  {\emph {\bibinfo {booktitle} {Advances in {Neural} {Information} {Processing}
  {Systems} 22}}},\ \bibinfo {editor} {edited by\ \bibinfo {editor}
  {\bibfnamefont {Y.}~\bibnamefont {Bengio}}, \bibinfo {editor} {\bibfnamefont
  {D.}~\bibnamefont {Schuurmans}}, \bibinfo {editor} {\bibfnamefont {J.~D.}\
  \bibnamefont {Lafferty}}, \bibinfo {editor} {\bibfnamefont {C.~K.~I.}\
  \bibnamefont {Williams}}, \ and\ \bibinfo {editor} {\bibfnamefont
  {A.}~\bibnamefont {Culotta}}}\ (\bibinfo  {publisher} {Curran Associates,
  Inc.},\ \bibinfo {year} {2009})\ pp.\ \bibinfo {pages}
  {1303--1311}\BibitemShut {NoStop}%
\bibitem [{\citenamefont {Höfling}\ and\ \citenamefont
  {Tibshirani}(2009)}]{hofling_estimation_2009}%
  \BibitemOpen
  \bibfield  {author} {\bibinfo {author} {\bibfnamefont {Holger}\ \bibnamefont
  {Höfling}}\ and\ \bibinfo {author} {\bibfnamefont {Robert}\ \bibnamefont
  {Tibshirani}},\ }\bibfield  {title} {\enquote {\bibinfo {title} {Estimation
  of sparse binary pairwise markov networks using pseudo-likelihoods},}\
  }\href@noop {} {\bibfield  {journal} {\bibinfo  {journal} {Journal of Machine
  Learning Research}\ }\textbf {\bibinfo {volume} {10}},\ \bibinfo {pages}
  {883--906} (\bibinfo {year} {2009})}\BibitemShut {NoStop}%
\bibitem [{\citenamefont {Nguyen}\ \emph {et~al.}(2017)\citenamefont {Nguyen},
  \citenamefont {Zecchina},\ and\ \citenamefont {Berg}}]{nguyen_inverse_2017}%
  \BibitemOpen
  \bibfield  {author} {\bibinfo {author} {\bibfnamefont {H.~Chau}\ \bibnamefont
  {Nguyen}}, \bibinfo {author} {\bibfnamefont {Riccardo}\ \bibnamefont
  {Zecchina}}, \ and\ \bibinfo {author} {\bibfnamefont {Johannes}\ \bibnamefont
  {Berg}},\ }\bibfield  {title} {\enquote {\bibinfo {title} {Inverse
  statistical problems: from the inverse {Ising} problem to data science},}\
  }\href {\doibase 10.1080/00018732.2017.1341604} {\bibfield  {journal}
  {\bibinfo  {journal} {Advances in Physics}\ }\textbf {\bibinfo {volume}
  {66}},\ \bibinfo {pages} {197--261} (\bibinfo {year} {2017})}\BibitemShut
  {NoStop}%
\bibitem [{\citenamefont {Gomez~Rodriguez}\ \emph {et~al.}(2010)\citenamefont
  {Gomez~Rodriguez}, \citenamefont {Leskovec},\ and\ \citenamefont
  {Krause}}]{gomez_rodriguez_inferring_2010}%
  \BibitemOpen
  \bibfield  {author} {\bibinfo {author} {\bibfnamefont {Manuel}\ \bibnamefont
  {Gomez~Rodriguez}}, \bibinfo {author} {\bibfnamefont {Jure}\ \bibnamefont
  {Leskovec}}, \ and\ \bibinfo {author} {\bibfnamefont {Andreas}\ \bibnamefont
  {Krause}},\ }\bibfield  {title} {\enquote {\bibinfo {title} {Inferring
  {Networks} of {Diffusion} and {Influence}},}\ }in\ \href {\doibase
  10.1145/1835804.1835933} {\emph {\bibinfo {booktitle} {Proceedings of the
  16th {ACM} {SIGKDD} {International} {Conference} on {Knowledge} {Discovery}
  and {Data} {Mining}}}},\ \bibinfo {series and number} {{KDD} '10}\ (\bibinfo
  {publisher} {ACM},\ \bibinfo {address} {New York, NY, USA},\ \bibinfo {year}
  {2010})\ pp.\ \bibinfo {pages} {1019--1028}\BibitemShut {NoStop}%
\bibitem [{\citenamefont {Myers}\ and\ \citenamefont
  {Leskovec}(2010)}]{myers_convexity_2010}%
  \BibitemOpen
  \bibfield  {author} {\bibinfo {author} {\bibfnamefont {Seth}\ \bibnamefont
  {Myers}}\ and\ \bibinfo {author} {\bibfnamefont {Jure}\ \bibnamefont
  {Leskovec}},\ }\bibfield  {title} {\enquote {\bibinfo {title} {On the
  {Convexity} of {Latent} {Social} {Network} {Inference}},}\ }in\ \href
  {http://papers.nips.cc/paper/4113-on-the-convexity-of-latent-social-network-inference.pdf}
  {\emph {\bibinfo {booktitle} {Advances in {Neural} {Information} {Processing}
  {Systems} 23}}},\ \bibinfo {editor} {edited by\ \bibinfo {editor}
  {\bibfnamefont {J.~D.}\ \bibnamefont {Lafferty}}, \bibinfo {editor}
  {\bibfnamefont {C.~K.~I.}\ \bibnamefont {Williams}}, \bibinfo {editor}
  {\bibfnamefont {J.}~\bibnamefont {Shawe-Taylor}}, \bibinfo {editor}
  {\bibfnamefont {R.~S.}\ \bibnamefont {Zemel}}, \ and\ \bibinfo {editor}
  {\bibfnamefont {A.}~\bibnamefont {Culotta}}}\ (\bibinfo  {publisher} {Curran
  Associates, Inc.},\ \bibinfo {year} {2010})\ pp.\ \bibinfo {pages}
  {1741--1749}\BibitemShut {NoStop}%
\bibitem [{\citenamefont {Netrapalli}\ and\ \citenamefont
  {Sanghavi}(2012)}]{netrapalli_learning_2012}%
  \BibitemOpen
  \bibfield  {author} {\bibinfo {author} {\bibfnamefont {Praneeth}\
  \bibnamefont {Netrapalli}}\ and\ \bibinfo {author} {\bibfnamefont {Sujay}\
  \bibnamefont {Sanghavi}},\ }\bibfield  {title} {\enquote {\bibinfo {title}
  {Learning the {Graph} of {Epidemic} {Cascades}},}\ }in\ \href {\doibase
  10.1145/2254756.2254783} {\emph {\bibinfo {booktitle} {Proceedings of the
  12th {ACM} {SIGMETRICS}/{PERFORMANCE} {Joint} {International} {Conference} on
  {Measurement} and {Modeling} of {Computer} {Systems}}}},\ \bibinfo {series
  and number} {{SIGMETRICS} '12}\ (\bibinfo  {publisher} {ACM},\ \bibinfo
  {address} {New York, NY, USA},\ \bibinfo {year} {2012})\ pp.\ \bibinfo
  {pages} {211--222}\BibitemShut {NoStop}%
\bibitem [{\citenamefont {Ma}\ \emph {et~al.}(2018)\citenamefont {Ma},
  \citenamefont {Chen}, \citenamefont {Lai},\ and\ \citenamefont
  {Zhang}}]{ma_statistical_2018}%
  \BibitemOpen
  \bibfield  {author} {\bibinfo {author} {\bibfnamefont {Chuang}\ \bibnamefont
  {Ma}}, \bibinfo {author} {\bibfnamefont {Han-Shuang}\ \bibnamefont {Chen}},
  \bibinfo {author} {\bibfnamefont {Ying-Cheng}\ \bibnamefont {Lai}}, \ and\
  \bibinfo {author} {\bibfnamefont {Hai-Feng}\ \bibnamefont {Zhang}},\
  }\bibfield  {title} {\enquote {\bibinfo {title} {Statistical inference
  approach to structural reconstruction of complex networks from binary time
  series},}\ }\href {\doibase 10.1103/PhysRevE.97.022301} {\bibfield  {journal}
  {\bibinfo  {journal} {Physical Review E}\ }\textbf {\bibinfo {volume} {97}},\
  \bibinfo {pages} {022301} (\bibinfo {year} {2018})}\BibitemShut {NoStop}%
\bibitem [{\citenamefont {Prasse}\ and\ \citenamefont
  {Van~Mieghem}(2018)}]{prasse_maximum-likelihood_2018}%
  \BibitemOpen
  \bibfield  {author} {\bibinfo {author} {\bibfnamefont {Bastian}\ \bibnamefont
  {Prasse}}\ and\ \bibinfo {author} {\bibfnamefont {Piet}\ \bibnamefont
  {Van~Mieghem}},\ }\bibfield  {title} {\enquote {\bibinfo {title}
  {Maximum-{Likelihood} {Network} {Reconstruction} for {SIS} {Processes} is
  {NP}-{Hard}},}\ }\href {http://arxiv.org/abs/1807.08630} {\bibfield
  {journal} {\bibinfo  {journal} {arXiv:1807.08630 [physics]}\ } (\bibinfo
  {year} {2018})},\ \bibinfo {note} {arXiv: 1807.08630}\BibitemShut {NoStop}%
\bibitem [{\citenamefont {{Braunstein Alfredo}}\ \emph
  {et~al.}(2019)\citenamefont {{Braunstein Alfredo}}, \citenamefont {{Ingrosso
  Alessandro}},\ and\ \citenamefont {{Muntoni Anna
  Paola}}}]{braunstein_alfredo_network_2019}%
  \BibitemOpen
  \bibfield  {author} {\bibinfo {author} {\bibnamefont {{Braunstein Alfredo}}},
  \bibinfo {author} {\bibnamefont {{Ingrosso Alessandro}}}, \ and\ \bibinfo
  {author} {\bibnamefont {{Muntoni Anna Paola}}},\ }\bibfield  {title}
  {\enquote {\bibinfo {title} {Network reconstruction from infection
  cascades},}\ }\href {\doibase 10.1098/rsif.2018.0844} {\bibfield  {journal}
  {\bibinfo  {journal} {Journal of The Royal Society Interface}\ }\textbf
  {\bibinfo {volume} {16}},\ \bibinfo {pages} {20180844} (\bibinfo {year}
  {2019})}\BibitemShut {NoStop}%
\bibitem [{\citenamefont {Runge}\ \emph {et~al.}(2012)\citenamefont {Runge},
  \citenamefont {Heitzig}, \citenamefont {Petoukhov},\ and\ \citenamefont
  {Kurths}}]{runge_escaping_2012}%
  \BibitemOpen
  \bibfield  {author} {\bibinfo {author} {\bibfnamefont {Jakob}\ \bibnamefont
  {Runge}}, \bibinfo {author} {\bibfnamefont {Jobst}\ \bibnamefont {Heitzig}},
  \bibinfo {author} {\bibfnamefont {Vladimir}\ \bibnamefont {Petoukhov}}, \
  and\ \bibinfo {author} {\bibfnamefont {Jürgen}\ \bibnamefont {Kurths}},\
  }\bibfield  {title} {\enquote {\bibinfo {title} {Escaping the {Curse} of
  {Dimensionality} in {Estimating} {Multivariate} {Transfer} {Entropy}},}\
  }\href {\doibase 10.1103/PhysRevLett.108.258701} {\bibfield  {journal}
  {\bibinfo  {journal} {Physical Review Letters}\ }\textbf {\bibinfo {volume}
  {108}},\ \bibinfo {pages} {258701} (\bibinfo {year} {2012})}\BibitemShut
  {NoStop}%
\bibitem [{\citenamefont {Sun}\ \emph {et~al.}(2015)\citenamefont {Sun},
  \citenamefont {Taylor},\ and\ \citenamefont {Bollt}}]{sun_causal_2015}%
  \BibitemOpen
  \bibfield  {author} {\bibinfo {author} {\bibfnamefont {Jie}\ \bibnamefont
  {Sun}}, \bibinfo {author} {\bibfnamefont {Dane}\ \bibnamefont {Taylor}}, \
  and\ \bibinfo {author} {\bibfnamefont {Erik~M.}\ \bibnamefont {Bollt}},\
  }\bibfield  {title} {{\selectlanguage {english}\enquote {\bibinfo {title}
  {Causal {Network} {Inference} by {Optimal} {Causation} {Entropy}},}\ }}\href
  {\doibase 10.1137/140956166} {\bibfield  {journal} {\bibinfo  {journal} {SIAM
  Journal on Applied Dynamical Systems}\ } (\bibinfo {year} {2015}),\
  10.1137/140956166}\BibitemShut {NoStop}%
\bibitem [{\citenamefont {Shen}\ \emph {et~al.}(2014)\citenamefont {Shen},
  \citenamefont {Wang}, \citenamefont {Fan}, \citenamefont {Di},\ and\
  \citenamefont {Lai}}]{shen_reconstructing_2014}%
  \BibitemOpen
  \bibfield  {author} {\bibinfo {author} {\bibfnamefont {Zhesi}\ \bibnamefont
  {Shen}}, \bibinfo {author} {\bibfnamefont {Wen-Xu}\ \bibnamefont {Wang}},
  \bibinfo {author} {\bibfnamefont {Ying}\ \bibnamefont {Fan}}, \bibinfo
  {author} {\bibfnamefont {Zengru}\ \bibnamefont {Di}}, \ and\ \bibinfo
  {author} {\bibfnamefont {Ying-Cheng}\ \bibnamefont {Lai}},\ }\bibfield
  {title} {{\selectlanguage {english}\enquote {\bibinfo {title} {Reconstructing
  propagation networks with natural diversity and identifying hidden
  sources},}\ }}\href {\doibase 10.1038/ncomms5323} {\bibfield  {journal}
  {\bibinfo  {journal} {Nature Communications}\ }\textbf {\bibinfo {volume}
  {5}},\ \bibinfo {pages} {4323} (\bibinfo {year} {2014})}\BibitemShut
  {NoStop}%
\bibitem [{\citenamefont {Ma}\ \emph {et~al.}(2015)\citenamefont {Ma},
  \citenamefont {Han}, \citenamefont {Shen}, \citenamefont {Wang},\ and\
  \citenamefont {Di}}]{ma_efficient_2015}%
  \BibitemOpen
  \bibfield  {author} {\bibinfo {author} {\bibfnamefont {Long}\ \bibnamefont
  {Ma}}, \bibinfo {author} {\bibfnamefont {Xiao}\ \bibnamefont {Han}}, \bibinfo
  {author} {\bibfnamefont {Zhesi}\ \bibnamefont {Shen}}, \bibinfo {author}
  {\bibfnamefont {Wen-Xu}\ \bibnamefont {Wang}}, \ and\ \bibinfo {author}
  {\bibfnamefont {Zengru}\ \bibnamefont {Di}},\ }\bibfield  {title}
  {{\selectlanguage {english}\enquote {\bibinfo {title} {Efficient
  {Reconstruction} of {Heterogeneous} {Networks} from {Time} {Series} via
  {Compressed} {Sensing}},}\ }}\href {\doibase 10.1371/journal.pone.0142837}
  {\bibfield  {journal} {\bibinfo  {journal} {PLOS ONE}\ }\textbf {\bibinfo
  {volume} {10}},\ \bibinfo {pages} {e0142837} (\bibinfo {year}
  {2015})}\BibitemShut {NoStop}%
\bibitem [{\citenamefont {Han}\ \emph {et~al.}(2015)\citenamefont {Han},
  \citenamefont {Shen}, \citenamefont {Wang},\ and\ \citenamefont
  {Di}}]{han_robust_2015}%
  \BibitemOpen
  \bibfield  {author} {\bibinfo {author} {\bibfnamefont {Xiao}\ \bibnamefont
  {Han}}, \bibinfo {author} {\bibfnamefont {Zhesi}\ \bibnamefont {Shen}},
  \bibinfo {author} {\bibfnamefont {Wen-Xu}\ \bibnamefont {Wang}}, \ and\
  \bibinfo {author} {\bibfnamefont {Zengru}\ \bibnamefont {Di}},\ }\bibfield
  {title} {\enquote {\bibinfo {title} {Robust {Reconstruction} of {Complex}
  {Networks} from {Sparse} {Data}},}\ }\href {\doibase
  10.1103/PhysRevLett.114.028701} {\bibfield  {journal} {\bibinfo  {journal}
  {Physical Review Letters}\ }\textbf {\bibinfo {volume} {114}},\ \bibinfo
  {pages} {028701} (\bibinfo {year} {2015})}\BibitemShut {NoStop}%
\bibitem [{\citenamefont {Li}\ \emph {et~al.}(2017)\citenamefont {Li},
  \citenamefont {Shen}, \citenamefont {Wang}, \citenamefont {Grebogi},\ and\
  \citenamefont {Lai}}]{li_universal_2017}%
  \BibitemOpen
  \bibfield  {author} {\bibinfo {author} {\bibfnamefont {Jingwen}\ \bibnamefont
  {Li}}, \bibinfo {author} {\bibfnamefont {Zhesi}\ \bibnamefont {Shen}},
  \bibinfo {author} {\bibfnamefont {Wen-Xu}\ \bibnamefont {Wang}}, \bibinfo
  {author} {\bibfnamefont {Celso}\ \bibnamefont {Grebogi}}, \ and\ \bibinfo
  {author} {\bibfnamefont {Ying-Cheng}\ \bibnamefont {Lai}},\ }\bibfield
  {title} {\enquote {\bibinfo {title} {Universal data-based method for
  reconstructing complex networks with binary-state dynamics},}\ }\href
  {\doibase 10.1103/PhysRevE.95.032303} {\bibfield  {journal} {\bibinfo
  {journal} {Physical Review E}\ }\textbf {\bibinfo {volume} {95}},\ \bibinfo
  {pages} {032303} (\bibinfo {year} {2017})}\BibitemShut {NoStop}%
\bibitem [{\citenamefont {Ching}\ \emph {et~al.}(2015)\citenamefont {Ching},
  \citenamefont {Lai},\ and\ \citenamefont
  {Leung}}]{ching_reconstructing_2015}%
  \BibitemOpen
  \bibfield  {author} {\bibinfo {author} {\bibfnamefont {Emily S.~C.}\
  \bibnamefont {Ching}}, \bibinfo {author} {\bibfnamefont {Pik-Yin}\
  \bibnamefont {Lai}}, \ and\ \bibinfo {author} {\bibfnamefont {C.~Y.}\
  \bibnamefont {Leung}},\ }\bibfield  {title} {\enquote {\bibinfo {title}
  {Reconstructing weighted networks from dynamics},}\ }\href {\doibase
  10.1103/PhysRevE.91.030801} {\bibfield  {journal} {\bibinfo  {journal}
  {Physical Review E}\ }\textbf {\bibinfo {volume} {91}},\ \bibinfo {pages}
  {030801} (\bibinfo {year} {2015})}\BibitemShut {NoStop}%
\bibitem [{\citenamefont {Lai}(2017)}]{lai_reconstructing_2017}%
  \BibitemOpen
  \bibfield  {author} {\bibinfo {author} {\bibfnamefont {Pik-Yin}\ \bibnamefont
  {Lai}},\ }\bibfield  {title} {\enquote {\bibinfo {title} {Reconstructing
  network topology and coupling strengths in directed networks of discrete-time
  dynamics},}\ }\href {\doibase 10.1103/PhysRevE.95.022311} {\bibfield
  {journal} {\bibinfo  {journal} {Physical Review E}\ }\textbf {\bibinfo
  {volume} {95}},\ \bibinfo {pages} {022311} (\bibinfo {year}
  {2017})}\BibitemShut {NoStop}%
\bibitem [{\citenamefont {Fortunato}\ and\ \citenamefont
  {Hric}(2016)}]{fortunato_community_2016}%
  \BibitemOpen
  \bibfield  {author} {\bibinfo {author} {\bibfnamefont {Santo}\ \bibnamefont
  {Fortunato}}\ and\ \bibinfo {author} {\bibfnamefont {Darko}\ \bibnamefont
  {Hric}},\ }\bibfield  {title} {\enquote {\bibinfo {title} {Community
  detection in networks: {A} user guide},}\ }\href {\doibase
  10.1016/j.physrep.2016.09.002} {\bibfield  {journal} {\bibinfo  {journal}
  {Physics Reports}\ } (\bibinfo {year} {2016}),\
  10.1016/j.physrep.2016.09.002}\BibitemShut {NoStop}%
\bibitem [{\citenamefont
  {Peixoto}(2017{\natexlab{a}})}]{peixoto_bayesian_2017}%
  \BibitemOpen
  \bibfield  {author} {\bibinfo {author} {\bibfnamefont {Tiago~P.}\
  \bibnamefont {Peixoto}},\ }\bibfield  {title} {\enquote {\bibinfo {title}
  {Bayesian stochastic blockmodeling},}\ }\href
  {http://arxiv.org/abs/1705.10225} {\bibfield  {journal} {\bibinfo  {journal}
  {arXiv:1705.10225 [cond-mat, physics:physics, stat]}\ } (\bibinfo {year}
  {2017}{\natexlab{a}})},\ \bibinfo {note} {arXiv: 1705.10225}\BibitemShut
  {NoStop}%
\bibitem [{\citenamefont {Berthet}\ \emph {et~al.}(2016)\citenamefont
  {Berthet}, \citenamefont {Rigollet},\ and\ \citenamefont
  {Srivastava}}]{berthet_exact_2016}%
  \BibitemOpen
  \bibfield  {author} {\bibinfo {author} {\bibfnamefont {Quentin}\ \bibnamefont
  {Berthet}}, \bibinfo {author} {\bibfnamefont {Philippe}\ \bibnamefont
  {Rigollet}}, \ and\ \bibinfo {author} {\bibfnamefont {Piyush}\ \bibnamefont
  {Srivastava}},\ }\bibfield  {title} {\enquote {\bibinfo {title} {Exact
  recovery in the {Ising} blockmodel},}\ }\href
  {http://arxiv.org/abs/1612.03880} {\bibfield  {journal} {\bibinfo  {journal}
  {arXiv:1612.03880 [math, stat]}\ } (\bibinfo {year} {2016})},\ \bibinfo
  {note} {arXiv: 1612.03880}\BibitemShut {NoStop}%
\bibitem [{\citenamefont {Hoffmann}\ \emph {et~al.}(2018)\citenamefont
  {Hoffmann}, \citenamefont {Peel}, \citenamefont {Lambiotte},\ and\
  \citenamefont {Jones}}]{hoffmann_community_2018}%
  \BibitemOpen
  \bibfield  {author} {\bibinfo {author} {\bibfnamefont {Till}\ \bibnamefont
  {Hoffmann}}, \bibinfo {author} {\bibfnamefont {Leto}\ \bibnamefont {Peel}},
  \bibinfo {author} {\bibfnamefont {Renaud}\ \bibnamefont {Lambiotte}}, \ and\
  \bibinfo {author} {\bibfnamefont {Nick~S.}\ \bibnamefont {Jones}},\
  }\bibfield  {title} {\enquote {\bibinfo {title} {Community detection in
  networks with unobserved edges},}\ }\href {http://arxiv.org/abs/1808.06079}
  {\bibfield  {journal} {\bibinfo  {journal} {arXiv:1808.06079 [physics]}\ }
  (\bibinfo {year} {2018})},\ \bibinfo {note} {arXiv: 1808.06079}\BibitemShut
  {NoStop}%
\bibitem [{\citenamefont {Newman}(2018)}]{newman_network_2018-1}%
  \BibitemOpen
  \bibfield  {author} {\bibinfo {author} {\bibfnamefont {M.~E.~J.}\
  \bibnamefont {Newman}},\ }\bibfield  {title} {{\selectlanguage
  {english}\enquote {\bibinfo {title} {Network structure from rich but noisy
  data},}\ }}\href {\doibase 10.1038/s41567-018-0076-1} {\bibfield  {journal}
  {\bibinfo  {journal} {Nature Physics}\ }\textbf {\bibinfo {volume} {14}},\
  \bibinfo {pages} {542--545} (\bibinfo {year} {2018})}\BibitemShut {NoStop}%
\bibitem [{\citenamefont {Peixoto}(2018)}]{peixoto_reconstructing_2018}%
  \BibitemOpen
  \bibfield  {author} {\bibinfo {author} {\bibfnamefont {Tiago~P.}\
  \bibnamefont {Peixoto}},\ }\bibfield  {title} {\enquote {\bibinfo {title}
  {Reconstructing {Networks} with {Unknown} and {Heterogeneous} {Errors}},}\
  }\href {\doibase 10.1103/PhysRevX.8.041011} {\bibfield  {journal} {\bibinfo
  {journal} {Physical Review X}\ }\textbf {\bibinfo {volume} {8}},\ \bibinfo
  {pages} {041011} (\bibinfo {year} {2018})}\BibitemShut {NoStop}%
\bibitem [{\citenamefont {Karrer}\ and\ \citenamefont
  {Newman}(2011)}]{karrer_stochastic_2011}%
  \BibitemOpen
  \bibfield  {author} {\bibinfo {author} {\bibfnamefont {Brian}\ \bibnamefont
  {Karrer}}\ and\ \bibinfo {author} {\bibfnamefont {M.~E.~J.}\ \bibnamefont
  {Newman}},\ }\bibfield  {title} {\enquote {\bibinfo {title} {Stochastic
  blockmodels and community structure in networks},}\ }\href {\doibase
  10.1103/PhysRevE.83.016107} {\bibfield  {journal} {\bibinfo  {journal}
  {Physical Review E}\ }\textbf {\bibinfo {volume} {83}},\ \bibinfo {pages}
  {016107} (\bibinfo {year} {2011})}\BibitemShut {NoStop}%
\bibitem [{\citenamefont
  {Peixoto}(2017{\natexlab{b}})}]{peixoto_nonparametric_2017}%
  \BibitemOpen
  \bibfield  {author} {\bibinfo {author} {\bibfnamefont {Tiago~P.}\
  \bibnamefont {Peixoto}},\ }\bibfield  {title} {\enquote {\bibinfo {title}
  {Nonparametric {Bayesian} inference of the microcanonical stochastic block
  model},}\ }\href {\doibase 10.1103/PhysRevE.95.012317} {\bibfield  {journal}
  {\bibinfo  {journal} {Physical Review E}\ }\textbf {\bibinfo {volume} {95}},\
  \bibinfo {pages} {012317} (\bibinfo {year} {2017}{\natexlab{b}})}\BibitemShut
  {NoStop}%
\bibitem [{\citenamefont
  {Peixoto}(2014{\natexlab{a}})}]{peixoto_efficient_2014}%
  \BibitemOpen
  \bibfield  {author} {\bibinfo {author} {\bibfnamefont {Tiago~P.}\
  \bibnamefont {Peixoto}},\ }\bibfield  {title} {\enquote {\bibinfo {title}
  {Efficient {Monte} {Carlo} and greedy heuristic for the inference of
  stochastic block models},}\ }\href {\doibase 10.1103/PhysRevE.89.012804}
  {\bibfield  {journal} {\bibinfo  {journal} {Physical Review E}\ }\textbf
  {\bibinfo {volume} {89}},\ \bibinfo {pages} {012804} (\bibinfo {year}
  {2014}{\natexlab{a}})}\BibitemShut {NoStop}%
\bibitem [{\citenamefont {Guimerà}\ and\ \citenamefont
  {Sales-Pardo}(2009)}]{guimera_missing_2009}%
  \BibitemOpen
  \bibfield  {author} {\bibinfo {author} {\bibfnamefont {Roger}\ \bibnamefont
  {Guimerà}}\ and\ \bibinfo {author} {\bibfnamefont {Marta}\ \bibnamefont
  {Sales-Pardo}},\ }\bibfield  {title} {\enquote {\bibinfo {title} {Missing and
  spurious interactions and the reconstruction of complex networks},}\ }\href
  {\doibase 10.1073/pnas.0908366106} {\bibfield  {journal} {\bibinfo  {journal}
  {Proceedings of the National Academy of Sciences}\ }\textbf {\bibinfo
  {volume} {106}},\ \bibinfo {pages} {22073 --22078} (\bibinfo {year}
  {2009})}\BibitemShut {NoStop}%
\bibitem [{\citenamefont {Pastor-Satorras}\ \emph {et~al.}(2015)\citenamefont
  {Pastor-Satorras}, \citenamefont {Castellano}, \citenamefont {Van~Mieghem},\
  and\ \citenamefont {Vespignani}}]{pastor-satorras_epidemic_2015}%
  \BibitemOpen
  \bibfield  {author} {\bibinfo {author} {\bibfnamefont {Romualdo}\
  \bibnamefont {Pastor-Satorras}}, \bibinfo {author} {\bibfnamefont {Claudio}\
  \bibnamefont {Castellano}}, \bibinfo {author} {\bibfnamefont {Piet}\
  \bibnamefont {Van~Mieghem}}, \ and\ \bibinfo {author} {\bibfnamefont
  {Alessandro}\ \bibnamefont {Vespignani}},\ }\bibfield  {title} {\enquote
  {\bibinfo {title} {Epidemic processes in complex networks},}\ }\href
  {\doibase 10.1103/RevModPhys.87.925} {\bibfield  {journal} {\bibinfo
  {journal} {Reviews of Modern Physics}\ }\textbf {\bibinfo {volume} {87}},\
  \bibinfo {pages} {925--979} (\bibinfo {year} {2015})}\BibitemShut {NoStop}%
\bibitem [{\citenamefont {Besag}(1974)}]{besag_spatial_1974}%
  \BibitemOpen
  \bibfield  {author} {\bibinfo {author} {\bibfnamefont {Julian}\ \bibnamefont
  {Besag}},\ }\bibfield  {title} {{\selectlanguage {english}\enquote {\bibinfo
  {title} {Spatial {Interaction} and the {Statistical} {Analysis} of {Lattice}
  {Systems}},}\ }}\href {\doibase 10.1111/j.2517-6161.1974.tb00999.x}
  {\bibfield  {journal} {\bibinfo  {journal} {Journal of the Royal Statistical
  Society: Series B (Methodological)}\ }\textbf {\bibinfo {volume} {36}},\
  \bibinfo {pages} {192--225} (\bibinfo {year} {1974})}\BibitemShut {NoStop}%
\bibitem [{\citenamefont {Metropolis}\ \emph {et~al.}(1953)\citenamefont
  {Metropolis}, \citenamefont {Rosenbluth}, \citenamefont {Rosenbluth},
  \citenamefont {Teller},\ and\ \citenamefont
  {Teller}}]{metropolis_equation_1953}%
  \BibitemOpen
  \bibfield  {author} {\bibinfo {author} {\bibfnamefont {Nicholas}\
  \bibnamefont {Metropolis}}, \bibinfo {author} {\bibfnamefont {Arianna~W.}\
  \bibnamefont {Rosenbluth}}, \bibinfo {author} {\bibfnamefont {Marshall~N.}\
  \bibnamefont {Rosenbluth}}, \bibinfo {author} {\bibfnamefont {Augusta~H.}\
  \bibnamefont {Teller}}, \ and\ \bibinfo {author} {\bibfnamefont {Edward}\
  \bibnamefont {Teller}},\ }\bibfield  {title} {\enquote {\bibinfo {title}
  {Equation of {State} {Calculations} by {Fast} {Computing} {Machines}},}\
  }\href {\doibase 10.1063/1.1699114} {\bibfield  {journal} {\bibinfo
  {journal} {The Journal of Chemical Physics}\ }\textbf {\bibinfo {volume}
  {21}},\ \bibinfo {pages} {1087} (\bibinfo {year} {1953})}\BibitemShut
  {NoStop}%
\bibitem [{\citenamefont {Hastings}(1970)}]{hastings_monte_1970}%
  \BibitemOpen
  \bibfield  {author} {\bibinfo {author} {\bibfnamefont {W.~K.}\ \bibnamefont
  {Hastings}},\ }\bibfield  {title} {\enquote {\bibinfo {title} {Monte {Carlo}
  sampling methods using {Markov} chains and their applications},}\ }\href
  {\doibase 10.1093/biomet/57.1.97} {\bibfield  {journal} {\bibinfo  {journal}
  {Biometrika}\ }\textbf {\bibinfo {volume} {57}},\ \bibinfo {pages} {97 --109}
  (\bibinfo {year} {1970})}\BibitemShut {NoStop}%
\bibitem [{\citenamefont {Decelle}\ \emph {et~al.}(2011)\citenamefont
  {Decelle}, \citenamefont {Krzakala}, \citenamefont {Moore},\ and\
  \citenamefont {Zdeborová}}]{decelle_asymptotic_2011}%
  \BibitemOpen
  \bibfield  {author} {\bibinfo {author} {\bibfnamefont {Aurelien}\
  \bibnamefont {Decelle}}, \bibinfo {author} {\bibfnamefont {Florent}\
  \bibnamefont {Krzakala}}, \bibinfo {author} {\bibfnamefont {Cristopher}\
  \bibnamefont {Moore}}, \ and\ \bibinfo {author} {\bibfnamefont {Lenka}\
  \bibnamefont {Zdeborová}},\ }\bibfield  {title} {\enquote {\bibinfo {title}
  {Asymptotic analysis of the stochastic block model for modular networks and
  its algorithmic applications},}\ }\href {\doibase 10.1103/PhysRevE.84.066106}
  {\bibfield  {journal} {\bibinfo  {journal} {Physical Review E}\ }\textbf
  {\bibinfo {volume} {84}},\ \bibinfo {pages} {066106} (\bibinfo {year}
  {2011})}\BibitemShut {NoStop}%
\bibitem [{Note1()}]{Note1}%
  \BibitemOpen
  \bibinfo {note} {Retrieved from \protect \url {openflights.org}.}\BibitemShut
  {Stop}%
\bibitem [{\citenamefont {Martinez}(1991)}]{martinez_artifacts_1991}%
  \BibitemOpen
  \bibfield  {author} {\bibinfo {author} {\bibfnamefont {Neo~D.}\ \bibnamefont
  {Martinez}},\ }\bibfield  {title} {{\selectlanguage {english}\enquote
  {\bibinfo {title} {Artifacts or {Attributes}? {Effects} of {Resolution} on
  the {Little} {Rock} {Lake} {Food} {Web}},}\ }}\href {\doibase
  10.2307/2937047} {\bibfield  {journal} {\bibinfo  {journal} {Ecological
  Monographs}\ }\textbf {\bibinfo {volume} {61}},\ \bibinfo {pages} {367--392}
  (\bibinfo {year} {1991})}\BibitemShut {NoStop}%
\bibitem [{Note2()}]{Note2}%
  \BibitemOpen
  \bibinfo {note} {Note that in this case our method also exploits the
  heterogeneous degrees in the network via the DC-SBM, which can in principle
  also aid the reconstruction, in addition to the community structure
  itself.}\BibitemShut {Stop}%
\bibitem [{Note3()}]{Note3}%
  \BibitemOpen
  \bibinfo {note} {Refinements of this approach including TAP and BP
  corrections~\cite {nguyen_inverse_2017} yield the same performance for this
  example.}\BibitemShut {Stop}%
\bibitem [{\citenamefont {Decelle}\ and\ \citenamefont
  {Ricci-Tersenghi}(2014)}]{decelle_pseudolikelihood_2014}%
  \BibitemOpen
  \bibfield  {author} {\bibinfo {author} {\bibfnamefont {Aurélien}\
  \bibnamefont {Decelle}}\ and\ \bibinfo {author} {\bibfnamefont {Federico}\
  \bibnamefont {Ricci-Tersenghi}},\ }\bibfield  {title} {\enquote {\bibinfo
  {title} {Pseudolikelihood {Decimation} {Algorithm} {Improving} the
  {Inference} of the {Interaction} {Network} in a {General} {Class} of {Ising}
  {Models}},}\ }\href {\doibase 10.1103/PhysRevLett.112.070603} {\bibfield
  {journal} {\bibinfo  {journal} {Physical Review Letters}\ }\textbf {\bibinfo
  {volume} {112}},\ \bibinfo {pages} {070603} (\bibinfo {year}
  {2014})}\BibitemShut {NoStop}%
\bibitem [{\citenamefont {Hodas}\ and\ \citenamefont
  {Lerman}(2014)}]{hodas_simple_2014}%
  \BibitemOpen
  \bibfield  {author} {\bibinfo {author} {\bibfnamefont {Nathan~O.}\
  \bibnamefont {Hodas}}\ and\ \bibinfo {author} {\bibfnamefont {Kristina}\
  \bibnamefont {Lerman}},\ }\bibfield  {title} {{\selectlanguage
  {english}\enquote {\bibinfo {title} {The {Simple} {Rules} of {Social}
  {Contagion}},}\ }}\href {\doibase 10.1038/srep04343} {\bibfield  {journal}
  {\bibinfo  {journal} {Scientific Reports}\ }\textbf {\bibinfo {volume} {4}},\
  \bibinfo {pages} {4343} (\bibinfo {year} {2014})}\BibitemShut {NoStop}%
\bibitem [{\citenamefont
  {Peixoto}(2014{\natexlab{b}})}]{peixoto_hierarchical_2014}%
  \BibitemOpen
  \bibfield  {author} {\bibinfo {author} {\bibfnamefont {Tiago~P.}\
  \bibnamefont {Peixoto}},\ }\bibfield  {title} {\enquote {\bibinfo {title}
  {Hierarchical {Block} {Structures} and {High}-{Resolution} {Model}
  {Selection} in {Large} {Networks}},}\ }\href {\doibase
  10.1103/PhysRevX.4.011047} {\bibfield  {journal} {\bibinfo  {journal}
  {Physical Review X}\ }\textbf {\bibinfo {volume} {4}},\ \bibinfo {pages}
  {011047} (\bibinfo {year} {2014}{\natexlab{b}})}\BibitemShut {NoStop}%
\bibitem [{\citenamefont {Holten}(2006)}]{holten_hierarchical_2006}%
  \BibitemOpen
  \bibfield  {author} {\bibinfo {author} {\bibfnamefont {D.}~\bibnamefont
  {Holten}},\ }\bibfield  {title} {\enquote {\bibinfo {title} {Hierarchical
  {Edge} {Bundles}: {Visualization} of {Adjacency} {Relations} in
  {Hierarchical} {Data}},}\ }\href {\doibase 10.1109/TVCG.2006.147} {\bibfield
  {journal} {\bibinfo  {journal} {IEEE Transactions on Visualization and
  Computer Graphics}\ }\textbf {\bibinfo {volume} {12}},\ \bibinfo {pages}
  {741--748} (\bibinfo {year} {2006})}\BibitemShut {NoStop}%
\bibitem [{\citenamefont
  {Peixoto}(2014{\natexlab{c}})}]{peixoto_graph-tool_2014}%
  \BibitemOpen
  \bibfield  {author} {\bibinfo {author} {\bibfnamefont {Tiago~P.}\
  \bibnamefont {Peixoto}},\ }\bibfield  {title} {\enquote {\bibinfo {title}
  {The \texttt{graph-tool} python library},}\ }\href {\doibase
  10.6084/m9.figshare.1164194} {\bibfield  {journal} {\bibinfo  {journal}
  {figshare}\ } (\bibinfo {year} {2014}{\natexlab{c}}),\
  10.6084/m9.figshare.1164194},\ \bibinfo {note} {available at
  \url{https://graph-tool.skewed.de}.}\BibitemShut {Stop}%
\end{thebibliography}%

\appendix

\section{Nonparametric DC-SBM model summary} \label{app:model}

The DC-SBM used in this work is the same derived in detail in
Ref.~\cite{peixoto_nonparametric_2017}.  We give a succinct summary in
the following. The marginal likelihood of the DC-SBM can be written as
\begin{align}\label{eq:dc-ensemble-equivalence}
  P(\A|\bb) &= \int P(\A|\bm\lambda,\bm\kappa,\bb)P(\bm\kappa|\bb)P(\bm\lambda|\bb)\;\dd\bm\kappa\,\dd\bm\lambda,\\
  &=
  P(\A|\bm{k},\e,\bb)P(\bm{k}|\e,\bb)P(\e|\bb),
\end{align}
where $\e=\{e_{rs}\}$ is the matrix of edge counts betwen groups, and
$\bm{k}$ is the degree sequence of the network, and with
\begin{align}
  P(\A|\bm{k},\e,\bb) &= \frac{\prod_{r<s}e_{rs}!\prod_re_{rr}!!\prod_ik_i!}{\prod_{i<j}A_{ij}!\prod_iA_{ii}!!\prod_re_r!!},\label{eq:micro-dc-sbm}\\
  P(\bm{k}|\e,\bb) &= \prod_r\multiset{n_r}{e_r}^{-1},\label{eq:micro-uniform-degrees}\\
  P(\e|\bb) &= \bar\lambda^E/(\bar\lambda+1)^{E+B(B+1)/2},
\end{align}
being the microcanonical likelihood and corresponding noninformative
priors. We further increase the explanatory power of this
model~\cite{peixoto_nonparametric_2017} by replacing the microcanonical
prior for the degrees with
\begin{equation}\label{eq:k_dist}
  P(\bm{k}|\bm{e},\bm{b}) = P(\bm{k}|\bm{\eta})P(\bm{\eta}|\bm{e},\bm{b})
\end{equation}
where $\bm{\eta}=\{\eta_k^r\}$ are the degree frequencies of each group,
with $\eta_k^r$ being the number of nodes with degree $k$ that belong to
group $r$, and
\begin{equation}
  P(\bm{k}|\bm{\eta}) = \prod_r \frac{\prod_k\eta_k^r!}{n_r!}
\end{equation}
is a uniform distribution of degree sequences constrained by the overall
degree counts, and finally
\begin{equation}\label{eq:k_dist_prior}
  P(\bm{\eta}|\bm{e},\bm{b}) = \prod_r q(e_r, n_r)^{-1}
\end{equation}
is the distribution of the overall degree counts. The quantity $q(m,n)$
is the number of different degree counts with the sum of degrees being
exactly $m$ and that have at most $n$ non-zero counts, given by
\begin{equation}\label{eq:q-exact}
  q(m, n) = q(m, n - 1) + q(m - n, n).
\end{equation}
For the node partition we use the prior,
\begin{equation}\label{eq:partition-prior}
  P(\bb) = P(\bb|\bm{n})P(\bm{n}|B)P(B)=\frac{\prod_rn_r!}{N!}{N-1\choose B-1}^{-1}N^{-1}.
\end{equation}
which is agnostic to group sizes.

Finally, the hierarchical degree-corrected SBM (HDC-SBM) is obtained by
replacing the uniform prior for $P(\e|\bb)$ by a nested sequence of
SBMs, where the edge counts in level $l$ are generated by a SBM at a
level above,
\begin{equation}\label{eq:multi_sbm}
  P(\bm{e}_l|\bm{e}_{l+1},\bm{b}_l) = \prod_{r<s}\multiset{n_r^ln^l_s}{e_{rs}^{l+1}}^{-1}
  \prod_{r}\multiset{n_r^l(n_r^l+1)/2}{e_{rr}^{l+1}/2}^{-1},
\end{equation}
where $\multiset{n}{m} = {n+m-1\choose m}$ is the multiset coefficient.
The prior for the hierarchical partition is obtained using
Eq.~\ref{eq:partition-prior} at every level. The entire model above is
also easily modified for directed networks. We refer to
Ref.~\cite{peixoto_nonparametric_2017} for further details.

\section{Adapting multigraph models to simple graphs}\label{app:multigraphs}

The DC-SBM variations considered above generate multigraphs with
self-loops, however the functional models presented in the main text
operate on simple graphs. We amend this inconsistency in the same manner
as in Ref.~\cite{peixoto_reconstructing_2018}, by adapting the
multigraph models to simple graphs in tractable way by generating
multigraphs and then collapsing the multiple edges. In other words, if
$\G$ is a multigraph with entries $G_{ij} \in \mathbb{N}$, the collapsed
simple graph $\A(\G)$ has binary entries
\begin{equation}
  A_{ij}(G_{ij}) =
  \begin{cases}
    1 & \text{ if } G_{ij} > 0 \text{ and } i\ne j,\\
    0 & \text{ otherwise.}
  \end{cases}
\end{equation}
Therefore, if $\G$ is a multigraph generated by $P(\G|\theta)$, where
$\theta$ are arbitrary parameters, then the corresponding collapsed simple
graph $\A$ is generated by
\begin{align}
  P(\A | \theta) &= \sum_{\G}P(\A,\G|\theta),\\
  &= \sum_{\G}P(\A|\G)P(\G|\theta),
\end{align}
with
\begin{equation}
  P(\A|\G) =
  \begin{cases}
    1 & \text{ if } \A = \A(\G),\\
    0 & \text{ otherwise.}
  \end{cases}
\end{equation}
Even if $P(\A | \theta)$ cannot be computed in closed form, the joint
distribution $P(\A,\G|\theta) = P(\A|\G)P(\G|\theta)$ is trivial,
provided we have $P(\G|\theta)$ in closed form. Therefore, instead of
directly sampling from the posterior distribution
\begin{equation}
  P(\A,\bb|\D) = \frac{P(\D|\A)P(\A,\bb)}{P(\D)},
\end{equation}
we sample from the joint posterior
\begin{equation}
  P(\A,\G,\bb|\D) = \frac{P(\D|\A)P(\A|\G)P(\G,\bb)}{P(\D)},
\end{equation}
using MCMC, treating the values $G_{ij}$ as latent variables, and then
we marginalize
\begin{equation}
  P(\A,\bb|\D) = \sum_{\G}P(\A,\G,\bb|\D),
\end{equation}
which is done simply by sampling from $P(\A,\G,\bb|\D)$ and ignoring the
actual magnitudes of the $G_{ij}$ values, and the diagonal entries.

\section{Inference algorithm}\label{app:algo}

The inference algorithm used here is identical to
Ref.~\cite{peixoto_reconstructing_2018}, with the only difference being
the likelihoods for the forward model $P(\D|\A)$. To summarize, we use
MCMC to sample from the joint posterior distribution
\begin{equation}\label{eq:joint_posterior_app}
  P(\A,\bb|\D) = \frac{P(\D|\A)P(\A|\bb)P(\bb)}{P(\D)},
\end{equation}
where $\bb$ is the partition of nodes used for the SBM.  The MCMC
algorithm consists of making proposals of the kind $P(\bb'|\A,\bb)$ and
$P(\A'|\A,\bb)$ for the partition and network, respectively (or
equivalently for any other remaining model parameter), and accepting
them according to the Metropolis-Hastings probability
\begin{equation}\label{eq:metropolis_app}
  \min\left(1, \frac{P(\A',\bb'|\D)P(\A|\A',\bb')P(\bb|\A',\bb')}{P(\A,\bb|\D)P(\A'|\A,\bb)P(\bb'|\A,\bb)}\right),
\end{equation}
which does not require the computation of the intractable normalization
constant $P(\D)$. In practice, at each step in the chain we make either
a move proposal for $\A$ or $\bb$, not both at once. For the node
partition, we use the move proposals described in
Refs.~\cite{peixoto_efficient_2014,peixoto_nonparametric_2017}, where
for any given node $i$ in group $r$ we propose to move it to group $s$
(which can be previously unoccupied, in which case it is labelled
$s=B+1$) according to
\begin{multline}
  P(b_i=r\to s|\A,\bb) = d\delta_{s,B+1} + {}\\
 (1-d)(1-\delta_{s,B+1})\sum_{t=1}^{B}P(t|i)\frac{e_{ts} + \epsilon}{e_t + \epsilon B},
\end{multline}
where $P(t|i) = \sum_jA_{ij}\delta_{b_j,t}/k_i$ is the fraction of
neighbours of $i$ that belong to group $t$, $\epsilon > 0$ is a small
parameter which guarantees ergodicity, and $d$ is the probability of
moving to a previously unoccupied group. (If $k_i = 0$, we assume
$P(b_i=r\to s|\A,\bb)=d\delta_{s,B+1} + (1-d)(1-\delta_{s,B+1})/B$.)
This move proposal attempts to the use the currently known large-scale
structure of the network to better inform the possible moves of the
node, without biasing with respect to group assortativity. The
parameters $d$ and $\epsilon$ do not affect the correctness of the
algorithm, only the mixing time, which is typically not very sensitive,
provided they are chosen within a reasonable range (we used $d=0.01$ and
$\epsilon=1$ throughout).  When using the HDC-SBM, we used the variation
of the above for hierarchical partitions described in
Ref.~\cite{peixoto_nonparametric_2017}. The move proposals above require
only minimal bookkeeping of the number edges incident on each group, and
can be made in time $O(k_i)$, which is also the time required to compute
the ratio in Eq.~\ref{eq:metropolis_app}, independent on how many groups
are currently occupied.

For the network, we change the values of the latent multigraph $\G$ with
unit proposals
\begin{equation}
  P(G_{ij}' = G_{ij} + \delta|\G) =
  \begin{cases}
    1/2 & \text{ if } G_{ij} > 0,\\
    1   & \text{ if } G_{ij} = 0 \text{ and } \delta = 1,\\
    0   & \text{ otherwise},
  \end{cases}
\end{equation}
for $\delta \in \{-1, 1\}$. We choose the entries to update with a probability given
by the current DC-SBM,
\begin{equation}\label{eq:move}
  P(i,j|\G,\bb) = \kappa_i\kappa_jm_{b_i,b_j},
\end{equation}
with
\begin{equation}
  \kappa_i = \frac{k_i+1}{\sum_j\delta_{b_j,b_i}k_j+1}
\end{equation}
being the probability of selecting node $i$ from its group $b_j$,
proportional to its current degree plus one, and
\begin{equation}
  m_{rs} = \frac{e_{rs} + 1}{\sum_{tu}e_{rs}+1}
\end{equation}
is the probability of selecting groups $(r,s)$, where
$e_{rs}=\sum_{ij}G_{ij}\delta_{b_i,r}\delta_{b_j,s}$. The above
probabilities guarantee that every entry will be eventually sampled, but
it tends to probe denser regions more frequently, which we found to
typically lead to faster mixing times. This sampling can be done in time
$O(1)$, simply by keeping urns of vertices and edges according to the
group memberships. The time required
to compute the ratio in Eq.~\ref{eq:metropolis_app} is also $O(1)$ for
the DC-SBM and $O(L)$ for the HDC-SBM, where $L$ is the hierarchy depth,
again independent of the number of occupied groups.

\subsection{Algorithmic complexity}

When combining the move proposals defined above for the partition and
network, the time required to perform $N$ node move proposals and $E$
edge addition or removal proposals is $O(\avg{k}N + EM)$, where
$\avg{k}$ is the average degree, and $M$ is the number of samples per
node of the functional model (i.e. SIS or Ising). The $O(EM)$
contribution is seen by noting that the addition and removal of an edge
requires the re-computation of the likelihood $P(\D|\A)$ involving only
terms associated with each endpoint over all $M$ samples, each requiring
only $O(1)$ computations. For the SIS model we note that we need only to
keep track of the summary quantities $m_i(t)$ for each node, and update
them by adding or subtracting contributions for each added or removed
edge, and the same is true for the Ising model with respect to edge
contributions to the Hamiltonian. This linear complexity of sweeps
allows for the reconstruction of large networks.

For dynamical data where changes of the state of each node are
relatively rare (e.g. in a SI or SIR dynamics, a node changes its state
only once or twice, respectively, for the whole cascade), it is possible
to optimize the inference algorithm by listing for each node only its
initial state and the times it changes, together with the new state
values. In this way, the contribution to the likelihood
of a single node can be computed by only going through the times that
its neighbours or the node itself change state, instead of the whole
time-span of the dynamics. Therefore, the complexity for a whole MCMC
sweep changes to $O(\avg{k}N + Ea)$, where $a$ is the average number of
times a single node changes its state during the whole dynamics. For
very active dynamics we have $a = O(M)$, and hence this algorithm has
the same complexity as the version above, but for $a \ll O(M)$ it gives
noticeable speed-ups.

In addition to the algorithmic complexity of each sweep, the MCMC needs
time to converge to the target posterior distribution. This mixing time
depends not only on the structure of the network being reconstructed,
how easy it is to uncover it from the data, but also on how close the
algorithm is initiated to the target distribution. Because of this, it
not straightforward to estimate the general algorithmic complexity of
the mixing time. In our numerical experiments we found that both
starting from random or empty networks lead to reasonable mixing times
in most cases, and the results coincide with initializing from the true
planted network (which as expected, shows faster equilibration).

A reference
implementation of the above algorithm is freely available as part of the
\texttt{graph-tool} library~\cite{peixoto_graph-tool_2014}.

\section{Datasets with empirical dynamics}\label{app:empirical}

\begin{figure}
  \includegraphics[width=\columnwidth]{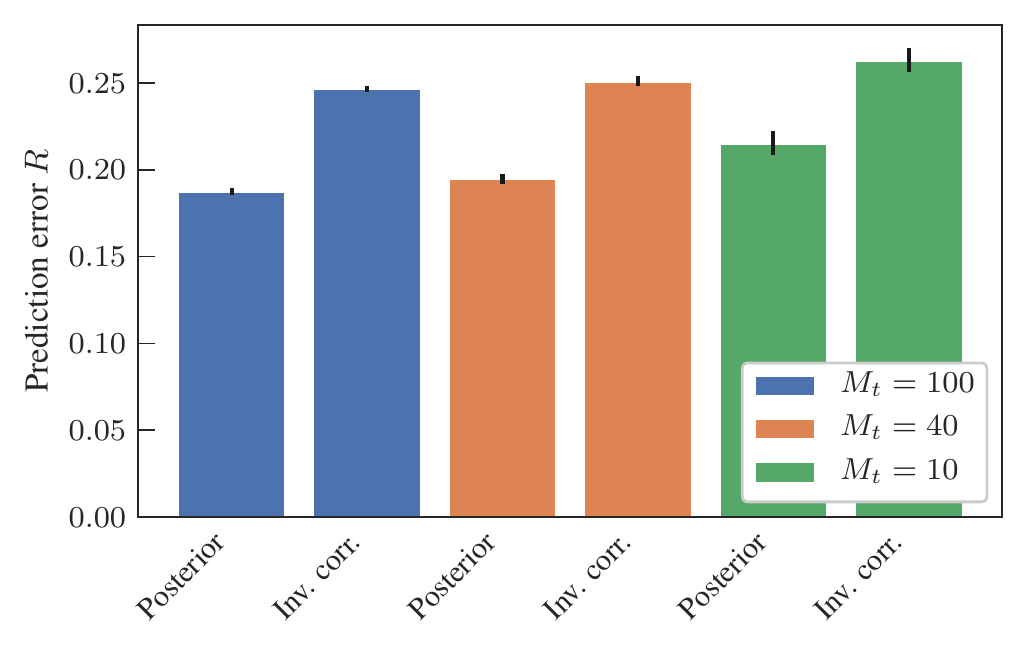}
  \caption{Cross-validation results for the  voting behavior of deputies in the lower house of the
Brazilian congress, comparing the posterior sampling approach considered
in the main text and the inverse correlation
method.\label{fig:votes-cross}}
\end{figure}

\begin{table}
  \begin{tabular}{l|p{23em}}
    Group & \multicolumn{1}{c}{User countries} \\ \hline\hline
1&$\text{Brazil}\times 216$, $\text{Japan}\times 2$, $\text{Germany}\times 2$, $\text{USA}\times 1$, $\text{Russia}\times 1$, $\text{Argentina}\times 1$, $\text{Colombia}\times 1$, $\text{France}\times 1$, $\text{South Africa}\times 1$\\\hline
2&$\text{Japan}\times 104$, $\text{UK}\times 1$, $\text{China}\times 1$, $\text{Italy}\times 1$, $\text{USA}\times 1$\\\hline
3&$\text{Indonesia}\times 42$, $\text{India}\times 10$, $\text{UK}\times 8$, $\text{Germany}\times 6$, $\text{USA}\times 6$, $\text{Thailand}\times 1$, $\text{Brazil}\times 1$, $\text{Bali}\times 1$, $\text{Argentina}\times 1$, $\text{Australia}\times 1$\\\hline
4&$\text{Indonesia}\times 52$, $\text{Germany}\times 6$, $\text{UK}\times 4$, $\text{USA}\times 3$, $\text{Russia}\times 2$, $\text{Australia}\times 2$, $\text{New Zealand}\times 1$, $\text{India}\times 1$, $\text{Botswana}\times 1$, $\text{Brazil}\times 1$\\\hline
5&$\text{USA}\times 54$, $\text{UK}\times 4$, $\text{Russia}\times 1$, $\text{Japan}\times 1$, $\text{Indonesia}\times 1$\\\hline
6&$\text{USA}\times 23$, $\text{Brazil}\times 9$, $\text{UK}\times 6$, $\text{Netherlands}\times 3$, $\text{Russia}\times 2$, $\text{Canada}\times 2$, $\text{Italy}\times 2$, $\text{Germany}\times 2$, $\text{Mexico}\times 1$, $\text{South Africa}\times 1$, $\text{Kwait}\times 1$, $\text{Austria}\times 1$, $\text{Romania}\times 1$, $\text{Finland}\times 1$, $\text{Japan}\times 1$, $\text{Philippines}\times 1$, $\text{Egypt}\times 1$, $\text{Argentina}\times 1$, $\text{Chile}\times 1$\\\hline
7&$\text{USA}\times 18$, $\text{India}\times 8$, $\text{Portugal}\times 2$, $\text{Brazil}\times 2$, $\text{Japan}\times 2$, $\text{UK}\times 2$, $\text{Guernsey}\times 1$, $\text{Malaysia}\times 1$, $\text{Mexico}\times 1$, $\text{Australia}\times 1$, $\text{Russia}\times 1$, $\text{France}\times 1$, $\text{Vietnam}\times 1$, $\text{Spain}\times 1$, $\text{Venezuela}\times 1$, $\text{Philippines}\times 1$\\\hline
8&$\text{Japan}\times 35$\\\hline
9&$\text{Brazil}\times 31$, $\text{USA}\times 2$, $\text{UK}\times 1$\\\hline
10&$\text{Korea}\times 24$, $\text{USA}\times 1$, $\text{Argentina}\times 1$, $\text{Russia}\times 1$, $\text{Japan}\times 1$\\\hline
11&$\text{Indonesia}\times 22$, $\text{UK}\times 2$, $\text{Australia}\times 1$, $\text{India}\times 1$, $\text{USA}\times 1$\\\hline
12&$\text{USA}\times 11$, $\text{Japan}\times 2$, $\text{Germany}\times 2$, $\text{France}\times 1$, $\text{Korea}\times 1$, $\text{Thailand}\times 1$, $\text{Brazil}\times 1$, $\text{Chile}\times 1$, $\text{Indonesia}\times 1$\\\hline
13&$\text{USA}\times 4$, $\text{Indonesia}\times 4$, $\text{India}\times 4$, $\text{UK}\times 3$, $\text{Australia}\times 2$, $\text{Canada}\times 1$\\\hline
14&$\text{Japan}\times 8$, $\text{Venezuela}\times 6$, $\text{USA}\times 1$, $\text{Chile}\times 1$\\\hline
15&$\text{Indonesia}\times 14$, $\text{Japan}\times 1$\\\hline
16&$\text{Indonesia}\times 7$, $\text{USA}\times 3$, $\text{Turkey}\times 1$, $\text{Philippines}\times 1$, $\text{Brazil}\times 1$\\\hline
17&$\text{USA}\times 9$, $\text{Canada}\times 1$, $\text{France}\times 1$, $\text{Belgium}\times 1$\\\hline
18&$\text{Germany}\times 5$, $\text{USA}\times 1$, $\text{Russia}\times 1$, $\text{France}\times 1$\\\hline
19&$\text{Japan}\times 4$
  \end{tabular} \caption{Country memberships of twitter users, according
  to the groups inferred by the reconstruction method.\label{tab:twitter}}
\end{table}

\begin{table}
  \begin{tabular}{l|p{23em}}
    Group & \multicolumn{1}{c}{Parties} \\ \hline\hline
1&$\text{\color{red}PMDB}\times 83$, $\text{\color{red}PT}\times 70$, $\text{\color{red}PP}\times 38$, $\text{\color{red}PR}\times 36$, $\text{\color{red}PSB}\times 27$, $\text{\color{red}PDT}\times 22$, $\text{\color{red}PTB}\times 17$, $\text{\color{red}PV}\times 14$, $\text{\color{red}PCdoB}\times 12$, $\text{\color{red}PSC}\times 10$, $\text{\color{blue}DEM}\times 5$, $\text{\color{red}PMN}\times 5$, $\text{\color{red}PRB}\times 3$, $\text{\color{red}PSOL}\times 3$, $\text{\color{red}PHS}\times 2$, $\text{\color{red}PTdoB}\times 1$, $\text{\color{red}PTC}\times 1$\\\hline
2&$\text{\color{blue}PSDB}\times 54$, $\text{\color{blue}PPS}\times 8$, $\text{\color{blue}PFL}\times 2$\\\hline
3&$\text{\color{blue}DEM}\times 41$\\\hline
4&$\text{\color{blue}DEM}\times 8$, $\text{\color{red}PMDB}\times 6$, $\text{\color{blue}PPS}\times 4$, $\text{\color{red}PP}\times 2$, $\text{\color{red}PSB}\times 1$\\\hline
  \end{tabular} \caption{Party affiliation of deputies of the lower
  house of the Brazilian congress, according to the groups they were
  classified by the reconstruction method. Parties in red belong to the
  center-left government coalition, and in blue to the right-wing
  opposition.\label{tab:parties}}
\end{table}

\subsection{Cross-validation}

To further evaluate the ability of our reconstruction method to capture
the empirical voting behavior of deputies in the lower house of the
Brazilian congress, we randomly divided all $M=619$ voting sessions into
$M-M_t$ ``training'' sessions, used to fit the model, and $M_t$ test
sessions, used to compare with the predictions from the model. To
evaluate the predicition error, the correlation matrix $C_{ij} =
\avg{s_is_j} - \avg{s_i}\avg{s_j}$ was computed from the posterior
distribution of the fitted model via
\begin{align}
  \avg{s_is_j} &= \sum_{\A,\bb,\beta,\bm J, \bm h}s_is_jP(\bm s|\A,\beta,\bm J, \bm h) P(\A,\bb,\beta,\bm J, \bm h | \bar{\bm s}_t)\\
  \avg{s_i} &= \sum_{\A,\bb,\beta,\bm J, \bm h}s_iP(\bm s|\A,\beta,\bm J, \bm h) P(\A,\bb,\beta,\bm J, \bm h | \bar{\bm s}_t)
\end{align}
where $\bar{\bm s}_t$ is the training data, and compared with the
correlation matrix obtained for the test data $C_{ij}^{(t)} = \avg{s_is_j}_t -
\avg{s_i}_t\avg{s_j}_t$, via
\begin{equation}
  \avg{s_is_j}_t = \frac{1}{M_t}\sum_m s_i^ms_j^m, \quad
  \avg{s_i}_t = \frac{1}{M_t}\sum_m s_i^m,
\end{equation}
where the sums go over microstates of the test data. The prediction
error is then computed as
\begin{equation}
  R = \frac{1}{{N\choose 2}}\sum_{i<j} |C_{ij} - C_{ij}^{(t)}|.
\end{equation}
We repeated the calculation using $M_t\in\{10, 40, 100\}$, and for each
value of $M_t$ we averaged the results over $30$ random choices of the
test data. We also compared with the reconstruction obtained via inverse
correlations~\cite{nguyen_inverse_2017}. The results are shown in
Fig.~\ref{fig:votes-cross}. As can be seen, the Bayesian joint
reconstruction method outperforms the results based on inverse
correlations. The prediction error decreases with larger $M_t$, as in
this limit the fluctuations in the dynamics are averaged out.

\subsection{Comparison with metadata}\label{app:metadata}

Here we expand on the comparison of the community structure found both
for the Brazilian congress as well as the twitter data, with metadata
available in both cases.

In table~\ref{tab:parties} we list the party affiliations of the
deputies according to the groups they were classified by the method. The
largest group accounts for all left-wing parties as well as the center
parties belonging to the government collation, whereas groups 2 and 3
accounts for the right-wing opposition. Group 4 is composed of a small
number deputies who are members of both government and oppositions
parties, but vote independently.

In table~\ref{tab:twitter} is shown the country of each twitter user,
independently obtained via twitter's API (not contained in the original
dataset of Ref.~\cite{hodas_simple_2014}), according to each group
identified by the reconstruction method. As can be seen, most groups are
characterized by a single dominating country, with only a few
exceptions. This indicates, plausibly, that the probability of re-tweets
is largely shaped by language and cultural barriers. Nevertheless, the
method also uncovers subdivisions within the distinct geographical
locations, indicating that this is not the only factor determining the
influence among users.

\end{document}